\documentclass[aps,amsmath,amssymb,showpacs,showkeys,longbibliography]{revtex4-2}
\usepackage[dvips]{graphicx}
\usepackage{times}
\usepackage{braket}
\usepackage{xcolor}
\usepackage{orcidlink}
\usepackage{hyperref}
\hypersetup{
  colorlinks=true,
  urlcolor=blue,
  linkcolor=red,
  citecolor=blue
}
\usepackage{float}
\usepackage{multirow}
\usepackage[sort&compress]{natbib}
\usepackage[shortcuts]{extdash}

\begin{document}
\title{Perturbative Renormalisation Group Improved Black Hole Solution and its Quasinormal Modes}

\author{Rupam Jyoti Borah\orcidlink{0009-0005-5134-0421}}
\email[Email: ]{rupamjyotiborah856@gmail.com}

\author{Umananda Dev Goswami\orcidlink{0000-0003-0012-7549}}
\email[Email: ]{umananda@dibru.ac.in}

\affiliation{Department of Physics, Dibrugarh University, Dibrugarh 786004, 
Assam, India}

\begin{abstract}
In this work, we construct a perturbative black hole (BH) solution motivated by renormalization group (RG) improvement and investigate the quasinormal modes (QNMs) of the BH under scalar field perturbations in both Schwarzschild--de Sitter (SdS) and Schwarzschild--anti--de Sitter (SAdS) backgrounds. To compute the QNMs in the SdS spacetime, we employ the 6th-order Padé-averaged WKB approximation method, while for the SAdS background we utilize the direct shooting method. We examine the dependence of the QNM frequencies on the free parameter of the solution. Furthermore, we analyze the time evolution of a scalar field perturbation around the BH and present the corresponding time-domain profiles. The QNMs are also extracted from the time-domain data using the matrix pencil method. Using the extracted QNM frequencies, we reconstruct the waveform and compare it with the original time-domain profile, finding good agreement between the two. The QNM frequencies obtained from the 6th-order Padé-averaged WKB method and the time-domain analysis in the SdS background, as well as those obtained from the direct shooting method and time-domain analysis in the SAdS spacetime, show very good consistency.
\end{abstract}


\maketitle                                                                      

\section{Introduction} \label{sec1}
General Relativity (GR) has remained as the most successful classical 
description of gravity since its formulation. Over the years, it has undergone 
rigorous experimental tests with astrophysical observations for its 
predictions, and has been consistently verified to remarkable accuracy. Two 
major achievements in recent times further highlight its success: the direct 
detection of gravitational waves (GWs) by the LIGO collaboration 
\cite{L1,L2,L3,L4,L5}, and the first-ever and subsequent images of BHs 
captured by the Event Horizon Telescope (EHT) collaboration 
\cite{E1,E2,E3,E4,E5,E6}. Despite these successes, GR has shown some 
limitations. For instance, it does not provide a satisfactory explanation for 
the observed accelerated expansion of the Universe (the dark energy problem), 
and it can't fully account for the missing mass problem, commonly called the 
dark matter problem \cite{F1,F2,F3,F4}. Moreover, when gravity is treated as a 
fundamental interaction at extremely high energy scales, where quantum effects 
are expected to play a crucial role, GR becomes inadequate. Thus, GR fails 
to explain the behaviour of gravity at very high energy scales where Quantum 
Gravity (QG) effects become dominant. While the Standard Model (SM) of 
particle physics successfully describes the strong, electromagnetic, and weak 
interactions within the framework of Quantum Field Theory (QFT), it fails 
to include gravity because GR, in its conventional form, is not renormalizable. 
The goal to unify GR with the SM remains one of the central challenges in 
theoretical physics. In pursuit of this goal, several candidate theories of 
QG have been proposed to describe gravity at very high energy scales 
\cite{QG1,QG2,QG3,QG4,QG5,QG6}. Loop Quantum Gravity (LQG) \cite{RLQ1,RLQ2,
RLQ3,RLQ4} and String Theory \cite{RST1,RST2,RST3,RST4,RST5} are among the 
most extensively studied approaches. String Theory aims to provide a unified 
description of all four fundamental interactions within a single theoretical 
framework, and it predicts the existence of the graviton as the mediator 
particle of gravitational interaction. In contrast, LQG focuses specifically 
on the quantization of the gravitational field; it suggests that spacetime 
possesses a discrete structure composed of fundamental quantum units.

Although LQG and String Theory are among the most extensively studied 
approaches to QG, their predictions have not yet been tested experimentally. 
The main difficulty lies in the extremely high energy scales at which their 
distinctive quantum gravitational effects are expected to appear. These 
energies are far beyond the capabilities of present day experimental 
facilities, making direct verification of these theories currently 
impractical. Due to these limitations, other theoretical approaches have been 
explored to investigate possible QG effects within accessible energy scales. 
One such approach is the framework of Effective Field Theory (EFT). In this 
perspective, GR is treated as an effective theory that remains valid up to a 
certain cutoff scale. This allows quantum corrections to be systematically 
incorporated into the theory, while ensuring that the theory reduces to 
classical GR in the low-energy regime. Several works have been successfully 
done by considering gravity as an EFT \cite{GREFT1,GREFT2,GREFT3,GREFT4,GREFT5,
GREFT6}. It is worth noting that EFT typically predicts only small quantum 
corrections at low energies. However, these corrections can become important 
in regions of extreme gravity, such as near the event horizon of a BH or in 
the early Universe. Therefore, BHs provide one of the most promising 
environments for probing QG effects within our current theoretical and 
observational reach. Motivated by this, many studies have investigated BH 
properties within the EFT framework \cite{M1,M2,M3,M4,M5,M6,M7,M8,M9}.

Apart from the approaches discussed above, another promising framework for 
QG is the asymptotically safe gravity \cite{AS1,AS2,AS3,AS4,AS5,AS6}. Within 
the framework of the asymptotic safety scenario of QG, the primary objective 
is to establish a consistent and predictive ultraviolet (UV) completeness of 
gravity. The central idea is that gravitational interactions may approach a 
non-Gaussian fixed point under the renormalization group (RG) flow at high 
energy scales. If such a fixed point exists, the theory remains finite and 
well defined in the UV regime. To investigate this possibility, the asymptotic 
safety approach employs a covariant formulation of the Wilsonian 
renormalization group \cite{RGN1,RGN2,RGN3}. Within this framework, Newton's 
gravitational constant $G$ and the cosmological constant $\Lambda$ are treated 
as scale dependent quantities that acquire explicit spacetime dependence, 
denoted by $G(x)$ and $\Lambda(x)$, where $x\equiv x^{\mu}$. The introduction 
of the effective average action and its associated functional RG equation 
provides a concrete realization of this program \cite{RG01,RG02}. In this 
framework, a scale dependent effective action 
is constructed, which implements a Wilsonian RG flow on the space of all 
diffeomorphism invariant functionals of the metric. The UV consistency of the 
theory is then governed by the existence of a non-Gaussian fixed point of the 
RG flow, rather than by a direct quantization of classical GR. Such effective 
average action is closely related to the standard effective action and 
defines a family of scale dependent EFTs labeled by a particular momentum 
(energy) scale. For a physical system characterized by a single momentum scale, 
quantum corrections can be incorporated by evaluating the effective action at 
the corresponding scale, thereby implementing renormalization group 
improvement. This approach thus provides a systematic framework for encoding 
QG effects through scale-dependent couplings. Motivated by these developments, 
Refs.~\cite{RG1,RG2} constructed an RG improved effective gravitational action and the 
corresponding modified field equations. In this framework, the running of the 
gravitational couplings obtained from the RG flow is incorporated into the 
classical action, leading to scale dependent corrections to Einstein’s field 
equations. The resulting field equations can be solved perturbatively by 
expanding around a classical GR background in low-energy limits. The 
parameters of the perturbative expansion quantify the deviation of the 
running couplings from their classical constant values. Such solutions capture 
the leading QG corrections and provide small but systematic modifications to 
BH spacetimes. The imprints of these corrections may be reflected in various 
BH observables, including quasinormal modes (QNMs), BH shadows, and other 
phenomena associated with strong gravitational fields 
\cite{OB1,OB2,OB3,OB4,OB5,OB6,OB7}.

QNMs represent the characteristic damped oscillations of BHs and provide 
direct information about their fundamental parameters. The associated complex 
frequencies depend on intrinsic properties such as the mass, angular momentum, 
and charge of the BHs. Recent studies have explored the effects of 
various parameters arising in different modified theories of gravity on BH 
perturbations and QNMs. In particular, studies including Kalb-Ramond gravity \cite{SB1}, 
traceless conformal electrodynamics \cite{SB2}, Lorentz-violating models \cite{SB3}, dyon-like dilatonic BHs \cite{SB4}, 
and Einstein-power-Yang-Mills gravity \cite{SB5} have 
demonstrated how the QNM spectrum depends on the parameters of the underlying 
models. 
More recently, QNMs and related BH observables have also been 
investigated for BHs immersed in polytropic scalar field gas backgrounds 
\cite{SB6}, ModMax BHs in a quintessence background \cite{SB7}, and BHs 
endowed with modified Chaplygin gas \cite{SB8}, which further highlights the 
sensitivity of QNMs to matter distributions in the modified gravity framework.
These studies highlight the importance of QNMs as a powerful probe to 
detect deviations from GR and can open a window for signatures of beyond GR 
physics. Any modification to the underlying spacetime geometry, 
particularly those arising from QG effects, can induce measurable shifts in 
these frequencies. In several QG frameworks, including LQG and String Theory, 
corrections to a classical BH metric naturally lead to deviations in the 
QNM spectrum. Moreover, within the context of the AdS/CFT correspondence, QNMs 
in asymptotically anti-de Sitter spacetimes are related to the relaxation 
dynamics of strongly coupled QFTs \cite{ac1,ac2}. This correspondence 
highlights the broader significance of QNMs, suggesting that their study can 
shed light on the quantum aspects of gravitational systems. For these reasons, 
QNMs serve as powerful probes of possible QG imprints in BH spacetimes and 
have been extensively investigated across a wide range of gravitational 
theories in recent years \cite{QN1,QN2,QN3,QN4,QN5,QN6,QN7,QN8,QN9,QN10}.

Motivated by the above considerations, in this study, we construct a 
perturbative BH solution from the RG improved field equations presented in 
Refs.~\cite{RG1,RG2} and investigate the QNMs of the corresponding BH
solution.

The rest of the paper is organized as follows. In Section \ref{sec2}, we 
briefly review the RG improved action and the associated RG improved field 
equations. In Section \ref{sec3}, we derive a perturbative BH solution that 
captures the QG effects arising from RG improvement in the low-energy regime. 
In Section \ref{sec4}, we analyze the QNMs of the BH in both Schwarzschild-de 
Sitter (SdS) and Schwarzschild-Anti-de Sitter (SAdS) spacetimes for a scalar 
field perturbation. In Section \ref{sec5}, we study the time-domain profile of 
a scalar field perturbation around the BH and extract the QNMs from the 
corresponding time evolution. Finally, in Section \ref{sec6}, we summarize 
our results and present our conclusions.

\section{Renormalisation group improved field equations} \label{sec2}
As stated earlier, the RG improvement provides a systematic way to incorporate 
scale dependent quantum corrections into classical gravitational dynamics. The 
RG improved field equations considered in this work admit two equivalent 
derivations in the literature. In Ref.~\cite{RG1}, the field equations were 
obtained directly by imposing covariance and consistency conditions. 
Subsequently, in Ref.~\cite{RG2}, the authors constructed an effective action 
whose variation with respect to the metric reproduces the field equations 
derived in Ref.~\cite{RG1}. In this section, we first outline the key 
assumptions and essential steps underlying the derivation of the RG-improved 
field equations in Ref.~\cite{RG1}. We then summarize the construction of the 
corresponding effective action following the Lagrangian framework developed 
in Ref.~\cite{RG2}.

In Ref.~\cite{RG1}, the authors considered a spacetime dependent cosmological 
constant $\Lambda(x)$ and Newton's constant $G(x)$, so Einstein field 
equations with an energy-momentum tensor $T_{\mu\nu}$ can be written as
\begin{equation}
G_{\mu\nu} = -\,\Lambda(x) g_{\mu\nu} + 8 \pi\, G(x) T_{\mu\nu} \label{eq1}
\end{equation}
Now, the Bianchi identities suggest the consistency condition for these 
equations as 
\begin{equation}
8 \pi (G T_{\mu\nu})^{; \mu} = \Lambda_{; \nu}, \label{eq2}
\end{equation}
where semicolon (;) denotes the usual covariant differentiation.
However, in the presence of spacetime dependent gravitational couplings $G(x)$ 
and $\Lambda(x)$, the consistency condition Eq.~\eqref{eq2} is no longer 
identically satisfied. Moreover, in the vacuum, Eq.~\eqref{eq2} implies that 
the cosmological constant must be a constant. It therefore becomes evident that 
the Bianchi identities, $G_{\mu\nu}^{\;\;\;\;\;; \mu}=0$, cannot be satisfied 
without introducing additional contributions. Consequently, the field 
equations must be supplemented by extra covariant derivative terms involving 
the spacetime dependent couplings $G(x)$ and $\Lambda(x)$, as well as terms 
involving the energy-momentum tensor $T_{\mu\nu}$. To proceed further in 
direction with convenience, $G(x)$ and $\Lambda(x)$ can be parametrized as
\begin{equation}
G(x) = \bar{G} e^{\chi}, \quad \Lambda = \bar{\Lambda} e^{\xi}, \label{eq3}
\end{equation}
where $\chi(x)$ and $\xi(x)$ are prescribed spacetime dependent functions 
whose behaviour is fixed by the RG flow, and consequently, they do not satisfy 
any equations of motion. $\bar{G}$ and $\bar{\Lambda}$ are arbitrary 
reference values. This parametrization, given in Eq.~\eqref{eq3}, imposes the 
restriction that the spacetime dependent couplings $G(x)$ and $\Lambda(x)$ 
are not allowed to change sign along the RG flow. A further requirement is 
that all additional kinetic terms vanish in the limit where $G(x)$ and 
$\Lambda(x)$ are constant, thereby ensuring the recovery of the classical 
Einstein field equations. With this parametrization and the consideration that
the kinetic terms should have minimum one derivative of $\chi(x)$ and $\xi(x)$, the Einstein field equations with all possible such kinetic terms can be 
written as
\begin{equation}
G_{\mu\nu} = -\, \Lambda g_{\mu\nu} + \vartheta
_{\mu\nu} + \tilde{\vartheta}_{\mu\nu} + \mathcal{F} (\xi_{; \mu} \chi_{; \nu} + \xi_{; \nu} \chi_{; \mu}) + \mathcal{H} g_{\mu\nu} \xi^{; \rho} \chi_{; \rho} + 8 \pi G T_{\mu\nu}, \label{eq4}
\end{equation}
where $g_{\mu\nu}$ is the metric tensor and
\begin{equation}
\vartheta_{\mu\nu}
= \mathcal{A} \xi_{;\mu}\xi_{;\nu}
+ \mathcal{B} g_{\mu\nu} \xi^{;\rho}\xi_{;\rho}
+ \mathcal{C} \xi_{;\mu;\nu}
+ \mathcal{E} g_{\mu\nu} \Box\xi 
\end{equation} is the energy-momentum tensor containing all possible kinetic 
terms of $\xi(x)$ and 
\begin{equation}
\tilde{\vartheta}_{\mu\nu}
= \tilde{\mathcal{A}} \chi_{,\mu}\chi_{,\nu}
+ \tilde{\mathcal{B}} g_{\mu\nu} \chi^{;\rho}\chi_{;\rho}
+ \tilde{\mathcal{C}} \chi_{;\mu;\nu}
+ \tilde{\mathcal{E}} g_{\mu\nu} \Box\chi 
\end{equation} is the energy-momentum tensor containing all possible kinetic 
terms of $\chi(x)$. All coefficients $\mathcal{A}$, $\mathcal{B}$, 
$\mathcal{C}$, $\mathcal{E}$, $\tilde{\mathcal{A}}$, $\tilde{\mathcal{B}}$, 
$\tilde{\mathcal{C}}$, $\tilde{\mathcal{E}}$, as well as $\mathcal{F}$ and 
$\mathcal{H}$, are  functions of $\xi(x)$ and $\chi(x)$. After a few 
substitutions and simplifications, the Bianchi identities that arise from 
Eq.~\eqref{eq4} are
\begin{align}
\vartheta_{\mu\nu}^{;\mu}&
+ \tilde{\vartheta}_{\mu\nu}^{;\mu}
- \bar{\Lambda} e^{\xi} \xi_{;\nu}
+ \mathcal{F}' \xi^{;\mu}\xi_{;\mu}\chi_{;\nu}
+ \dot{\mathcal{F}} \chi^{;\mu}\chi_{;\mu}\xi_{;\nu}
+ (\mathcal{F}' + \mathcal{H}') \xi^{;\mu}\chi_{;\mu}\xi_{;\nu}
+ (\dot{\mathcal{F}} + \dot{\mathcal{H}}) \xi^{;\mu}\chi_{;\mu}\chi_{;\nu} 
\notag \\[5pt]
& + \mathcal{F}\, \Box\xi \chi_{;\nu}
+ \mathcal{F}\, \Box\chi \xi_{;\nu}
+ (\mathcal{F} + \mathcal{H}) \xi^{;\mu}\chi_{;\mu;\nu}
+ (\mathcal{F} + \mathcal{H}) \chi^{;\mu}\xi_{;\mu;\nu}
+ 8\pi \bar{G} e^{\chi}\!
\left( T_{\mu\nu}^{;\mu} + T_{\mu\nu}\chi^{;\mu} \right)
= 0,
\label{eq7}
\end{align}
where the prime and dot denote differentiation with respect to $\xi(x)$ and 
$\chi(x)$, respectively. The coefficients $\mathcal{A}$, $\mathcal{B}$, 
$\mathcal{C}$, $\mathcal{E}$, $\tilde{\mathcal{A}}$, $\tilde{\mathcal{B}}$, 
$\tilde{\mathcal{C}}$, $\tilde{\mathcal{E}}$ can be determine from the 
Bianchi relation~\eqref{eq7} as
\begin{equation}
\mathcal{A} = -\,\frac{1}{2}, \quad \mathcal{B} =-\, \frac{1}{4}, \quad \mathcal{C} = 1, \quad \mathcal{E} = -1, \quad \tilde{\mathcal{A}} = \tilde{\mathcal{B}} = \tilde{\mathcal{C}} = \tilde{\mathcal{E}} = \mathcal{H} = \mathcal{F} = 0.
\end{equation}
Thus, Eq.~\eqref{eq4} finally becomes:
\begin{equation}
G_{\mu\nu} = 8 \pi G T_{\mu\nu} - \bar{\Lambda}\, e^{\xi}\, g_{\mu\nu} - \frac{1}{2}\, \xi_{;\mu}\xi_{;\nu} -\frac{1}{4} g_{\mu\nu} \xi^{; \rho} \xi_{;\rho} + \xi_{;\mu ;\nu} - g_{\mu\nu} \Box \xi
\label{eq9}
\end{equation} 
with conservation equation:
\begin{equation}
\left(G T_{\mu\nu}\right)^{;\mu}
+ G \left( T_{\mu\nu} - \frac{1}{2}\, T g_{\mu\nu} \right)\xi^{;\mu} = 0.
\end{equation}

Motivated by the results of Ref.~\cite{RG1}, the authors of Ref.~\cite{RG2} 
developed a Lagrangian framework for the field equations given in 
Eq.~\eqref{eq9}. In fact, they constructed an effective action by 
treating the field equations as input, such that variation of this action with 
respect to the metric tensor reproduces Eq.~\eqref{eq9}. The steps involved 
in the construction of the action are as follows. As the field equations 
contain the Einstein tensor, which is not coupled to any additional fields, 
therefore, the Ricci scalar must be included in the action. Furthermore, the 
last two terms, $\xi_{;\mu ;\nu}$ and $g_{\mu\nu} \Box \xi$, appearing in 
Eq.~\eqref{eq9}, naturally arise in scalar-tensor theories due to the 
nonminimal coupling between the scalar field and gravity. These considerations 
indicate that the scalar field $\xi(x)$ must be coupled nonminimally to the 
Ricci tensor, while preserving the linear appearance of the Einstein tensor in 
the equations of motion. Thus, the Lagrangian should have a dilatonic type 
coupling and can be written in the following form \cite{RG2}:
\begin{equation}
\mathcal{L} \sim F(\xi)\bigl[R + \hdots\bigr],
\end{equation}
where $F(\xi)$ is a function of the scalar field $\xi(x)$. The 3rd 
and 4th terms in Eq.~\eqref{eq9} i.e. $\frac{1}{2}\, \xi_{;\mu}\xi_{;\nu}$ and 
$\frac{1}{4} g_{\mu\nu} \xi^{; \rho} \xi_{;\rho}$ can be obatined from a 
kinetic term of field $\xi(x)$ appearing in the Lagragian of the form:
\begin{equation}
\mathcal{L} \sim \omega (\xi_{;\mu})^2, 
\end{equation}
where $\omega$ is a coupling parameter. Further, the 1st term 
$8 \pi G T_{\mu\nu}$ emerges due to a nonminimal coupling of matter Lagrangian 
$\mathcal{L}_m$ with the Newton's constant and the 2nd term 
$\bar{\Lambda} e^{\xi} g_{\mu\nu}$, is the potential energy term $V(\xi)$. 
Moreover, since by construction $\xi(x)$ and $\chi(x)$ are not dynamical 
fields, variations of the action of the considered theory with respect to 
$\xi(x)$ and $\chi(x)$ are not required. Therefore, the required action of the 
theory will finally take the following form:
\begin{equation}
\mathcal{S} = \int d^4x \sqrt{-g} F(\xi) \left[R - \omega (\xi_{;\mu})^2 -V(\xi) + 16 \pi G(x) \mathcal{L}_m \right]. \label{eq14}
\end{equation}
The variation of this action \eqref{eq14} with respect to $g_{\mu\nu}$ gives
\begin{align}
G_{\mu\nu} = &\; 8\pi G T_{\mu\nu} - \frac{1}{2} V(\xi)\, g_{\mu\nu}
+ \left( \omega + \frac{F''}{F} \right) \xi_{;\mu} \xi_{;\nu} 
- g_{\mu\nu} \left( \frac{\omega}{2} + \frac{F''}{F} \right) (\xi_{;\mu})^2
\notag\\
& + \frac{F'}{F} \left(\xi_{;\mu ;\nu} - g_{\mu\nu}\Box\xi \right).
\label{eq15}
\end{align}
Substituting, $F(\xi) = e^{\xi}$, $\omega = -\,3/2$ and 
$V(x) = 2 \bar{\Lambda} e^{\xi}$ in Eq.~\eqref{eq15} and after a few 
simplifications we will recover the field Eq.~\eqref{eq9}. Since the field 
equations follow from a well defined action, the theory admits a consistent 
EFT interpretation. In the next section, we solve the field equations 
\eqref{eq9} for a Schwarzschild-(A)dS like BHs within this EFT framework. 

\section{Renormalisation Group improved Black Hole solution} \label{sec3}
In this section, we solve the field equations \eqref{eq9} in the vacuum to 
obtain a BH solution. In Eq.~\eqref{eq9}, $\xi(x)$ is a predefined spacetime 
function; therefore, its functional form depends on the underlying spacetime 
structure. For the Schwarzschild case, the Kretschmann scalar $K$ depends on 
the RG scale $\tilde{k}$ as follows \cite{Kres01,Kres02}:
\begin{equation}
K \propto \tilde{k}^4 \label{eq015}
\end{equation}
Again, 
\begin{equation}
K = R_{\mu\nu\sigma\rho} R^{\mu\nu\sigma\rho} = \frac{48 M^2}{r^6}, \label{eq016}
\end{equation}
where $M$ is the mass parameter and $R_{\mu\nu\sigma\rho}$ is the curvature 
tensor. Eq.~\eqref{eq015} tells us $\tilde{k} \propto K^{\frac{1}{4}}$ and using  
Eq.~\eqref{eq016} one can write
\begin{equation}
\tilde{k} \propto \left(\frac{48 M^2}{r^6}\right)^{\frac{1}{4}}
\propto \frac{M^{1/2}}{r^{3/2}}.
\end{equation}
This leads to
\begin{equation}
{\tilde{k}}^2 \propto \frac{M}{r^3} =  \frac{\zeta M}{r^3}, 
\label{eq20}
\end{equation}
where $\zeta$ is a proportionality constant. As the proportionality 
relation specifies only the $1/r^3$ dependence of the RG scale, the 
proportionality coefficient $\zeta$ parametrizes this dependence. Since 
$\zeta$ is unknown, it can be treated as a free parameter, which controls 
the strength of the RG-induced corrections. The mass parameter $M$ is kept 
explicit because the curvature of spacetime is produced by the BH mass. 
Moreover, for the Schwarzschild-(A)dS spacetime, the Kretschmann scalar 
contains both a mass dependent contribution and a cosmological constant 
contributions as
\begin{equation}
K = \frac{48 M^2}{r^6} + \frac{8}{3}{\Lambda}^2.
\end{equation}
Therefore, we decompose the RG scale into a constant asymptotic part, $k_{0}$, 
accounting for the $\Lambda^2$ term \cite{Kres02}, and a radial contribution 
sourced by the BH mass, as given in Eq.~\eqref{eq20}. Hence, the RG scale 
for the Schwarzschild-(A)dS spacetime takes form:
\begin{equation}
k^2 = k_{0}^2 + \frac{\zeta M}{r^3}. \label{eq22}
\end{equation}
It is evident that the cosmological constant $\Lambda(k)$ depends on the RG 
scale $k$ as $\Lambda(k) \propto k^2$. Comparing this scaling with the 
parametrization $\Lambda = \bar{\Lambda} e^{\xi}$ shows that $\xi(x)$ can be 
identified as
\begin{equation}
\xi = \ln k^2  + \ln (C\bar{\Lambda}^{-\,1}),
\end{equation}
where $C$ is a constant and can be absorbed into a redefinition of 
$\bar{\Lambda}$. Thus, $\xi(x)$ can be define as
\begin{equation}
\xi = \ln k^2. \label{eq24}
\end{equation}
Substituting Eq.~\eqref{eq22} in Eq.~\eqref{eq24}, $\xi(x)$ can be written as
\begin{equation}
\xi = \ln \left(k_{0}^2 + \frac{\zeta M}{r^3}\right). \label{eq25}
\end{equation}
Expanding Eq.~\eqref{eq25} in powers of $1/r$, one can find,
\begin{equation}
\xi =
\ln k_0^2
+\frac{\zeta M}{k_0^2 r^3}
-\frac{1}{2}\frac{\zeta^2 M^2}{k_0^4 r^6}
+O\!\left(\frac{\zeta^3 M^3}{k_0^6 r^9}\right). \label{eq26}
\end{equation}
Further, Eq.~\eqref{eq26} can be rewritten as
\begin{equation}
\xi = \xi_0 +\frac{\epsilon}{r^3} - \frac{1}{2}\frac{\epsilon^2}{r^6}
+O\!\left(\frac{\epsilon^3}{r^9}\right), \label{eq27}
\end{equation}
where $\xi_0 = \ln k_0^2$ and $\epsilon=\zeta M/k_0^2$. The parameter 
$\epsilon = \zeta M/k_0^2$ controls the strength of the RG induced corrections 
to the classical geometry. The perturbative solution obtained in this work is 
valid in the regime $\epsilon \ll 1$, for which the RG corrections represent 
a small deformation of the Schwarzschild-(A)dS spacetime. In this regime, 
higher-order terms in the expansion of $\xi$ and the metric functions are 
parametrically suppressed, ensuring the consistency of the perturbative 
approach. Consequently, the present analysis does not probe the strong 
quantum gravity regime near the singularity, but rather captures leading-order 
RG effects in the semiclassical domain. Since $\bar{\Lambda}$ is taken as a 
reference value of the cosmological constant in the EFT framework, we treat it 
as the cosmological constant in the IR limit and consider small values of 
$\bar{\Lambda}$ of order $10^{-3}$ in our analysis.

In this work, our aim is to construct a perturbative RG-improved BH 
solution with a minimal deformation from the classical Schwarzschild-(A)dS 
spacetime within the perturbative regime $\epsilon \ll 1$, because the 
RG-improved corrections are expected to remain small in this regime. So, we 
restrict our analysis to geometries that preserve the Schwarzschild-like 
structure order-by-order in the perturbative expansion. This choice also 
simplifies the modified field equations while still capturing the 
leading-order effects of the RG-improved corrections on the BH spacetime. 
Therefore, for a static and spherically symmetric spacetime with the above 
considerations, we can have the following ansatz:
\begin{equation}
ds^2 = -f(r)\,dt^2 + \frac{1}{f(r)}\,dr^2 + r^2 d\Omega^2.
\label{eq28}
\end{equation}
where ${d\Omega}^2 = d\theta^2 + \sin^2\!\theta\, d\phi^2$ . It needs to be 
mentioned that the choice $g_{tt} \neq -1/g_{rr}$ may lead to a more general 
class of RG-improved geometries with more complicated field equations and 
perturbative dynamics compared to the Schwarzschild-like form considered in 
the present work.

In Eq.~\eqref{eq28},
\begin{equation}
f(r) = f_0(r) + \sum\limits_{i\,=\,1}^{\infty}\epsilon^i a_i(r)
\end{equation} with
\begin{equation}
f_0(r) = 1 - \frac{2M}{r} - \frac{\bar{\Lambda}}{3}\, r^2. \label{eq29}
\end{equation}
Here, $f_0(r)$ represents the Schwarzschild-(A)dS solution of GR expressed in 
terms of the infrared (IR) 
cosmological constant $\bar{\Lambda}$, where $\bar{\Lambda}>0$ corresponds to 
the SdS case and $\bar{\Lambda}<0$ to the SAdS case. The second term in $f(r)$ 
represents perturbative corrections. The function $f(r)$ is expanded as a 
power series in the parameter $\epsilon$, with the coefficient functions 
$a_i(r)$ encoding successive higher-order deviations from the GR background. 
Now, our task is to determine the functions $a_i(r)$ by solving the field 
equations~\eqref{eq9} order by order in the perturbative expansion. 

For the first order correction $f(r)$ will have the form:
\begin{equation}
f(r) = f_0(r) + \epsilon a_1(r). \label{eq30}
\end{equation}
For vacuum, i.e., for $T_{\mu\nu} = 0$, Eq.~\eqref{eq9} reduces as
\begin{equation}
G_{\mu\nu} = -\, \bar{\Lambda} e^{\xi} g_{\mu\nu}
- \frac{1}{2}\, \xi_{;\mu} \xi_{;\nu}
- \frac{1}{4}\, g_{\mu\nu} (\xi_{;\mu})^2
+ \xi_{;\mu;\nu}
- g_{\mu\nu} \Box\xi , \label{eq31}
\end{equation}
where the Einstein tensor on the left-hand side (LHS) has the following forms 
for the $tt$, $rr$ and $\theta\theta$ components with the metric \eqref{eq28}:
\begin{align}
G_{tt} & = -f\, \frac{rf' + f - 1}{r^2}, \label{eq32}\\[5pt]
G_{rr} & = \,\frac{1}{f} \frac{rf' + f - 1}{r^2}, \label{eq33}\\[5pt]
G_{\theta\theta} & = \frac{r^2 f''}{2} + r f'. \label{eq34}
\end{align}
Here, primes denote the differentiation with respect to $r$. Using the 
definition of $\xi(x)$ in Eq.~\eqref{eq27} upto 1st-order in $\epsilon$, the 
different terms of the right-hand side (RHS) of Eq.~\eqref{eq31} for the 
$\theta\theta$ component can be written as follows:
\begin{align}
-\,\bar{\Lambda}e^{\xi}g_{\theta\theta} & = -\,\bar{\Lambda}e^{\xi_0} r^2 \left(1 + \frac{\epsilon}{r^3}\right),\\[5pt]
\xi_{;\theta;\theta} & = -\frac{3\epsilon f_0 }{r^3},\\[5pt]
-\,g_{\theta\theta}\Box{\xi} & = -\,\epsilon \left(\frac{6f_0}{r^3} - \frac{3{f'_0}}{r^2}\right).
\end{align}
Further, the second and third terms on the right-hand side of Eq.~\eqref{eq31}
involve the square of the gradient of $\xi$. Since the first derivative of 
$\xi$ is already of first order in the perturbative parameter $\epsilon$, 
these terms are of the $\mathcal{O}(\epsilon^2)$ and therefore do not 
contribute at first order in the perturbative expansion. Thus, for 
$\theta\theta$ component the RHS of Eq.~\eqref{eq31} takes the form:
\begin{equation}
\text{RHS}_{\theta\theta} = -\bar{\Lambda}e^{\xi_0}r^2 + \epsilon \left(-\,\frac{\bar{\Lambda}e^{\xi_0}}{r} - \frac{3 {f_0}}{r^3} + \frac{3{f'_0}}{r^2}\right). \label{eq38}
\end{equation}
Moreover, as $\bar{\Lambda}$ is an arbitrary reference value, the constant 
factor $e^{\xi_0}$ can be absorbed into the redefinition of $\bar{\Lambda}$. 
Hence, Eq.~\eqref{eq38} can be written as
\begin{equation}
\text{RHS}_{\theta\theta} = -\bar{\Lambda}r^2 + \epsilon \left(-\,\frac{\bar{\Lambda}}{r} - \frac{9 {f_0}}{r^3} + \frac{3{f'_0}}{r^2}\right). \label{eq39}
\end{equation}
Substituting Eq.~\eqref{eq30} in Eq.~\eqref{eq34}, the $G_{\theta\theta}$ 
component can be expanded upto 1st-order in $\epsilon$ as
\begin{equation}
G_{\theta\theta} = \frac{r^2 f''_0}{2} + r f'_0 + \epsilon \left(\frac{r^2}{2}\,a''_1 + r\,a'_1 \right).
\end{equation}
Finally, the $\theta\theta$ component of Eq.~\eqref{eq31} can be written as
\begin{equation}
\frac{r^2 f''_0}{2} + r f'_0  + \epsilon \left(\frac{r^2}{2}\,a''_1 + r a'_1 \right) =  -\bar{\Lambda}r^2 + \epsilon \left(-\,\frac{\bar{\Lambda}}{r} - \frac{9 {f_0}}{r^3} + \frac{3f'_0}{r^2} \right). \label{eq41}
\end{equation}
Equating the coefficients of $\epsilon$ on both sides of Eq.~\eqref{eq41}
yields the following 1st-order differential equation:
\begin{equation}
\frac{r^2}{2}\,a''_1 + r\,a'_1 = -\,\frac{\bar{\Lambda}}{r} - \frac{9 {f_0}}{r^3} + \frac{3{f'_0}}{r^2}. \label{eq42}
\end{equation}
Again, substitution of $f_0$ and $f'_0$ from Eq.~\eqref{eq29}, this 
Eq.~\eqref{eq42} reduces as
\begin{equation}
\frac{r^2}{2}\,a''_1 + r\,a'_1 = -\,\frac{9}{r^3} + \frac{24 M}{r^4}. \label{eq43}
\end{equation}
Solving this Eq.~\eqref{eq43}, $a_1$ can be obtain as
\begin{equation}
a_1(r) = -\, \frac{3}{r^3} + \frac{4 M}{r^4} -\, \frac{C_1}{r} + C_2,
\end{equation}
where $C_1$ and $C_2$ are constants of integration. In the asymptotic limit $r \to \infty$, the metric function behaves as
\begin{equation}
f(r) \to 1 - \frac{\bar{\Lambda}}{3} r^2 ,
\end{equation}
which corresponds to the standard asymptotic structure of an (A)dS spacetime. To ensure that the obtained solution preserves this asymptotic behavior, the integration constant $C_2$ must vanish. Further, the term $\frac{C_1}{r}$ contributes to the coefficient of the $1/r$ term in the metric function, which is directly associated with the ADM mass of the spacetime. Therefore, a non-zero value of $C_1$ would effectively lead to a redefinition of the mass parameter $M$. To maintain the physical interpretation of $M$ as the ADM mass of the BH, we impose the regularity condition $C_1 = 0$. Therefore, the RG improved BH 
solution upto 1st-order in $\epsilon$ is as follows:
\begin{equation}
f(r) =  1 - \frac{2M}{r} - \frac{\bar{\Lambda}}{3}\, r^2 + \epsilon \left(-\, \frac{3}{r^3} + \frac{4M}{r^4} \right). \label{eq45}
\end{equation}
%
In this solution, the additional inverse-power corrections arise from the 
RG-improved effective gravitational couplings and represent leading-order 
QG-induced modifications to the classical Schwarzschild-(A)dS geometry. These 
terms decay faster than the Schwarzschild contribution, and spacetime 
asymptotically approaches the classical GR solution at large distances. These 
correction terms produce deviations from the GR solution, which become 
significant in the near-horizon strong-field region. These corrections modify 
the effective potential governing perturbations and lead to shifts in the 
QNMs spectrum. Thus, Eq.~\eqref{eq45} represents a Schwarzschild-(A)dS BH 
metric function with leading order RG-improved perturbative correction.
Following the same process, the BH solution up to 2nd-order in $\epsilon$ can 
be obtained as
\begin{equation}
f(r) =  1 - \frac{2M}{r} - \frac{\bar{\Lambda}}{3}\, r^2 + \epsilon \left(-\, \frac{3}{r^3} + \frac{4M}{r^4} \right) + {\epsilon}^2 \left(\frac{21}{20r^6} - \frac{25M}{14r^7} - \frac{13 \bar{\Lambda}}{24r^4} \right). \label{eq46}
\end{equation}
Similarly, higher-order corrections can be systematically obtained by 
following the same procedure. Since the solution is perturbative, with 
$\epsilon \ll 1$, contributions from higher-order terms are increasingly 
suppressed. Hence, we limit our analysis to the second-order solution. 

It should be noted that the leading RG-improved correction to the metric 
function behaves as $\sim 1/r^{3}$, which resembles the radial falloff 
associated with quadrupolar contributions in multipole expansions of 
gravitational fields. However, a true quadrupole moment in GR is characterized 
by both radial contribution of the form $\propto 1/r^{3}$ and also by a 
nontrivial angular structure, which involves terms proportional to 
$P_{2}(\cos\theta)$ (Legendre polynomial), reflecting the breaking of 
spherical symmetry in the system \cite{QPM1,QPM2,QPM3,QPM4}. The present 
RG-improved solution remains spherically symmetric, since the metric function 
depends only on the radial coordinate $r$. Thus, the $1/r^{3}$ scaling term 
obtained here can not be interpreted as a true quadrupolar deformation of the 
spacetime geometry. Instead, it represents an isotropic correction arising 
from the RG running of the gravitational couplings through the scale 
identification $k^{2}\sim M/r^{3}$.

In all the numerical analyses presented in this work, 
the parameters $M$, $\zeta$, and $k_0$ are chosen such that the quantity 
$\epsilon = \zeta M / k_0^2$ remains $\ll 1$, ensuring the validity of the 
perturbative approach. Since $k_0$ represents the asymptotic value of the RG 
scale $k(r)$ in the limit $r \to \infty$, i.e. value of $k(r)$ in the IR 
energy scale and the precise numerical value of it is not fixed by the EFT, 
we must instead choose its value within a physically reasonable range. In 
geometrized units, values of $k_0 \sim 1 - 10$ correspond to very low 
energy scales in conventional units and therefore lie firmly in the IR regime. 
As the validity of the perturbative analysis requires the parameter 
$\epsilon \ll 1$, we fix $k_0 = 10$ in our analysis and consider both positive 
and negative values of $\zeta$, choosing its values such that the perturbative 
condition $\epsilon \ll 1$ is satisfied, and $k^2$ remains positive in 
Eq.~\eqref{eq22}. We consider both positive and negative values of $\zeta$ to 
study the dependence of the results on this parameter over a wider range.

\begin{figure}[!h]
\centerline{
\includegraphics[scale=0.87]{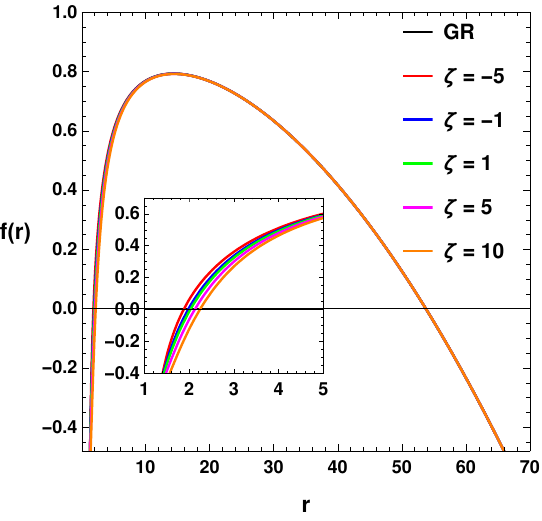} \hspace{0.7cm}
\includegraphics[scale=0.87]{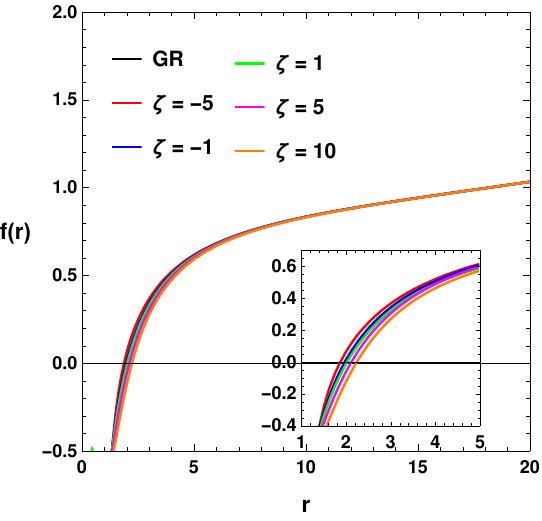}}
\vspace{-0.2cm}
\caption{Horizon structure of the BH solution \eqref{eq46} with respect to 
the parameter $\zeta$, obtained by fixing  $M = 1$, $\bar{\Lambda} = 0.001$ 
(left) and $\bar{\Lambda} = -0.001$ (right).} 
\label{fig01}
\end{figure}
Fig.~\ref{fig01} shows the horizon structure of the BH solution 
\eqref{eq46} for both SdS and SAdS cases. The left panel shows the horizon 
structure of the SdS spacetime for different values of $\zeta$. In this case,
there are two horizons of the BHs: the event horizon and the cosmological 
horizon. It can be observed from the figure that both negative and positive 
values of $\zeta$ produce noticeable shifts in the event horizon structure 
and negligible change in the cosmological horizon. The right panel 
corresponds to the horizon structure of the SAdS case, where there is only the 
event horizon. In this case, both positive and negative value of $\zeta$ 
changes the horizon structure of the solution.
\section{Quasinormal modes of Renormalisation Group improved black holes} \label{sec4}
In this section, we investigate the QNMs of the BH solution presented in 
Eq.~\eqref{eq46} for both SdS and SAdS spacetimes. The analysis is carried out 
for massless scalar field perturbations propagating in the background of the 
BH spacetime. 

For a massless scalar field $\Phi$ the equation of motion is given 
by \cite{Konoplya2023}
\begin{equation}
\frac{1}{\sqrt{-g}}\,\partial_{\mu}\!\left(\sqrt{-g}g^{\mu\nu}\partial_{\nu}\Phi \right) = 0. \label{eq47}
\end{equation} 
Assuming a spherical symmetry, the scalar field $\Phi$ can be decomposed into 
a radial part $R_{\omega,l}(r)$ and an angular part described by the spherical 
harmonics $Y_l(\theta,\phi)$, and can be expressed as
\begin{equation}
\Phi = e^{{\pm}\,i{\omega}t}\frac{{\Psi_{\omega,l}(r)}}{r}\,Y_{l}({\theta},{\phi}), 
\label{eq48}
\end{equation} 
where $l$ denotes the multipole number, $\omega$ is the frequency of 
oscillation, and $\Psi_{\omega,l}(r)$ can be defined as
\begin{equation}
\Psi_{\omega,l}(r) = r R_{{\omega},l}(r).
\end{equation} 
By substituting Eq.~\eqref{eq48} into Eq.~\eqref{eq47}, we obtain the 
following Schrödinger-like equation:
\begin{equation}
\frac{{d^2}\Psi}{d{r_{\star}}^2} + ({\omega}^2-V(r))\Psi=0, 
\label{eq50}
\end{equation}
where $r_{\star}$ denotes the tortoise coordinate, which is defined as
\begin{equation}
r_{\star} = \int{\frac{dr}{f(r)}},
\end{equation}
and $V(r)$ is the effective potential as given by
\begin{equation}
V(r) = f(r)\left(\frac{f'(r)}{r} + \frac{l(l+1)}{r^2}\right).
\end{equation}
For an asymptotically flat spacetime, Eq.~\eqref{eq50} must satisfy the 
following boundary conditions:
\begin{equation}
\Psi(r_{\star}) \rightarrow 
\begin{cases}
A e^{+i\omega r_*}, & \text{as } r_{\star} \rightarrow -\infty, \\[6pt]
B e^{-i\omega r_*}, & \text{as } r_{\star} \rightarrow +\infty .
\end{cases}
\end{equation}
Here, $A$ and $B$ represent the amplitudes of the respective wave components. 
These two conditions respectively correspond to purely ingoing waves at the 
BH event horizon and purely outgoing waves at spatial infinity, embodying the 
physical requirement that no information can escape from the horizon and no 
radiation can originate from infinity. The imposition of these boundary 
conditions results in an infinite, discrete set of complex frequencies, which 
are identified as the QNMs of the BH. 
\subsection{QNMs in Schwarzschild-de Sitter spacetime} \label{sub1}
The SdS spacetime describes a Universe with a positive cosmological constant. 
It features both a BH event horizon and a cosmological horizon 
\cite{sds1,sds2,sds3}. In this section, we study the QNMs of the BH described
by Eq.~\eqref{eq46} for a positive cosmological constant corresponding to the 
SdS spacetime. 

\begin{figure}[!h]
\centerline{
\includegraphics[scale=0.87]{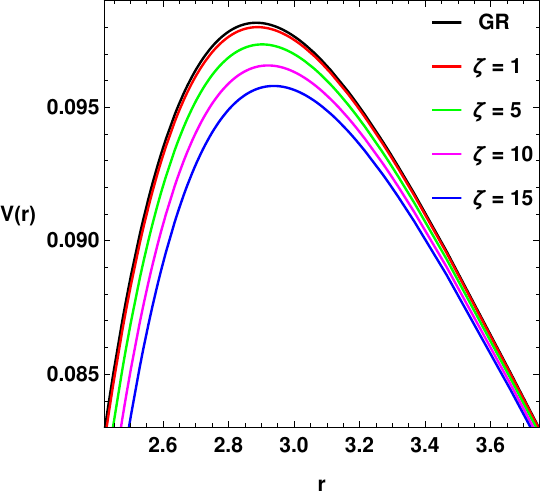} \hspace{0.7cm}
\includegraphics[scale=0.87]{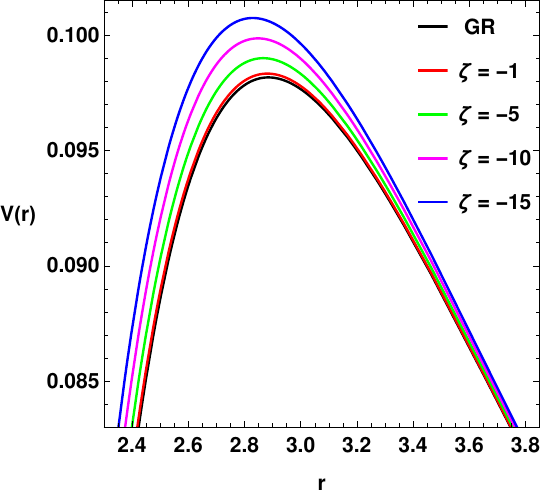}}
\vspace{-0.2cm}
\caption{Behaviour of the effective potential $V(r)$ as a function of the 
radial coordinate $r$ for both positive and negative values of the
parameter $\zeta$, obtained by fixing  $M = 1$ and $\bar{\Lambda} = 0.001$.} 
\label{fig1}
\end{figure}

Figure~\ref{fig1} illustrates the behaviour of the effective potential 
$V(r)$ as a function of the radial coordinate $r$ for different values of 
the parameter $\zeta$ in the SdS case. The left panel shows that positive 
values of $\zeta$ lower the peak of the potential compared to the GR case. 
This indicates that positive RG corrections lower the effective potential 
barrier experienced by perturbations. In contrast, the right panel illustrates 
that for negative values of $\zeta$, the peak of the potential is higher than 
the GR case, indicating stronger trapping of perturbative modes near the 
potential maximum. The dependence of $V(r)$ on $r$ for different magnitudes 
and signs of $\zeta$ indicates a clear sensitivity of the QNM spectrum to RG 
improved corrections.

We compute the QNMs using the 6th-order Pad\'e-averaged WKB approximation 
method, characterizing them in terms of their oscillation frequency and 
damping rate, and analyze their dependence on the parameter $\zeta$. 
Table~\ref{table1} presents the QNMs frequencies for different positive values 
of $\zeta$ and for multipole numbers $l = 1$, $2$, $3$. It can be observed 
from Table~\ref{table1} that both the oscillation frequency and the damping 
rate decrease as the parameter $\zeta$ increases for each value of $l$. The 
deviation from the GR case remains small, as expected for small perturbative 
quantum corrections. Further, for positive values of the parameter $\zeta$, the real part of the QNMs increases with $l$, while the imaginary part decreases as $l$ increases. Table~\ref{table2} displays the QNM frequencies for 
different negative values of the parameter $\zeta$ and for multipole numbers 
$l = 1$, $2$, $3$. From Table~\ref{table2}, it is evident that both the 
oscillation frequency and the damping rate increase as $\zeta$ is increased 
within the negative parameter regime for each value of $l$. Moreover, the dependence of the real and imaginary parts of the QNMs on $l$ for negative values of $\zeta$ is the same as that for positive values of $\zeta$. The error associated with the QNMs obtained from the 6th order Pad\'e averaged WKB method is calculated using the formula
\begin{equation}
\Delta_6 = \frac{|\text{Pad\'e averaged WKB}_7 - \text{Pad\'e averaged WKB}_5|}{2}
\end{equation}

\begin{figure}[!h]
    \centerline{
    \includegraphics[scale = 0.86]{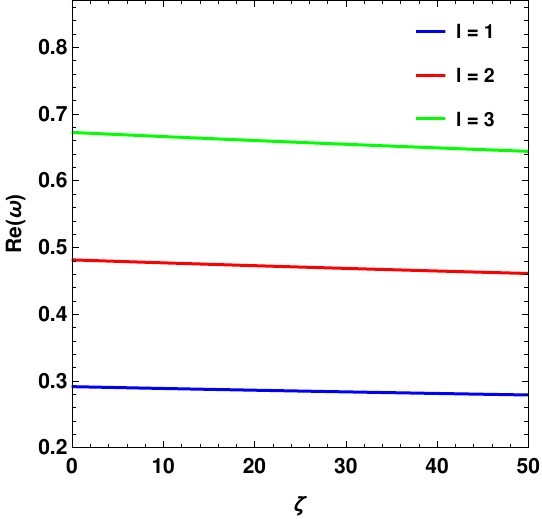}\hspace{0.5cm}
    \includegraphics[scale = 0.92]{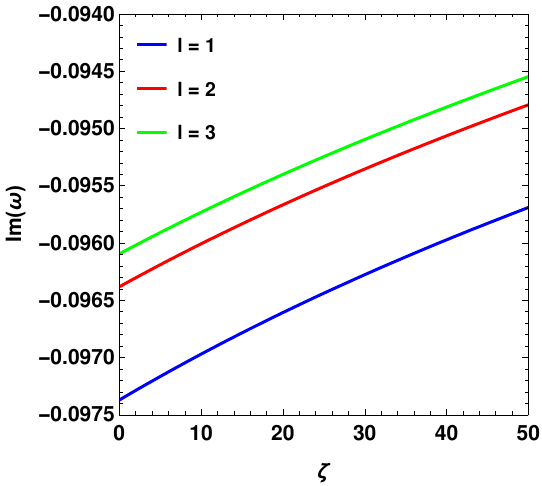}}
    \vspace{-0.2cm}
    \caption{Variation of real and imaginary parts of QNMs with respect 
to positive values of $\zeta$ for different values of $l$ along with 
$M = 1$, $n = 0$ and $\bar{\Lambda} = 0.001$.}
    \label{fig2}
\end{figure}

\begin{figure}[!h]
    \centerline{
    \includegraphics[scale = 0.86]{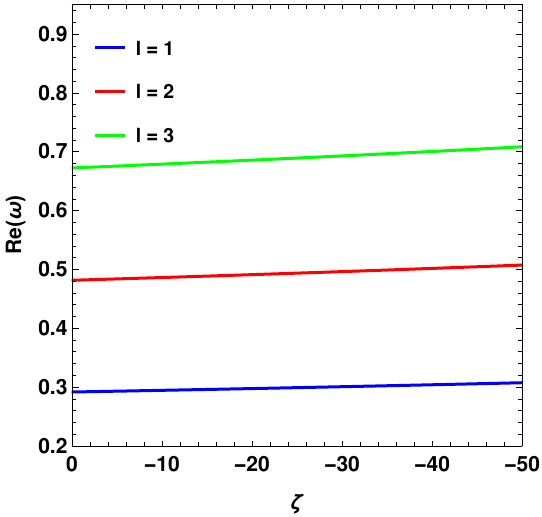}\hspace{0.5cm}
    \includegraphics[scale = 0.92]{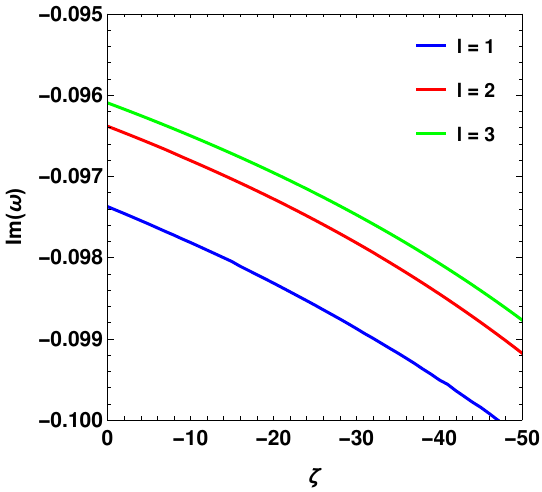}}
    \vspace{-0.2cm}
    \caption{Variation of real and imaginary parts of QNMs with respect 
to negative values of $\zeta$ for different values of $l$ along with 
$M = 1$, $n = 0$ and $\bar{\Lambda} = 0.001$.}
    \label{fig3}
\end{figure}

\begin{table}[!h]
\centering
\caption{QNMs of BHs described by the solution \eqref{eq46}, computed using 
the 6th-order Pad\'e-averaged WKB approximation method for different positive 
values of the parameter $\zeta$, with $M =1$, $n = 0$ and 
$\bar{\Lambda} = 0.001$.}
\vspace{5pt}
\begin{tabular}{c@{\hskip 5pt}c@{\hskip 10pt}c@{\hskip 10pt}c@{\hskip 10pt}c@{\hskip 5pt}c}
\hline\hline
& Multipole & $\zeta$ & Pad\'e averaged 6th-order WKB & $\Delta_{6}$ \\
\hline
& GR ($l=1$) & 0 & $0.291340 - 0.097472i$ & $0.3140 \times 10^{-5}$ & \\ \hline
&\multirow{4}{4em}{$l=1$} 
  & $1$ & $0.291058 - 0.097442i$ & $0.0261 \times 10^{-5}$ &\\
&  & $5$ & $0.289942 - 0.097328i$ & $0.0150 \times 10^{-5}$ &\\
&  & $10$ & $0.288574 - 0.097194i $ & $0.0124 \times 10^{-4}$ &\\
&  & $15$ & $0.287235 - 0.097067i$ & $0.3463 \times 10^{-6}$ &\\
\hline
& GR ($l=2$) & 0 & $0.481284 - 0.0963911i$ & $0.5298 \times 10^{-4}$ &\\ \hline
&\multirow{4}{4em}{$l=2$} 
  & $1$ & $0.480829 - 0.096352i$  & $0.0175 \times 10^{-4}$ &\\
&  & $5$ & $0.479032 - 0.096200i$  & $0.0256 \times 10^{-5}$ &\\
&  & $10$ & $0.476832 - 0.096018i$  & $0.0128 \times 10^{-6}$ &\\
&  & $15$ & $0.474682 - 0.095846i$  & $0.1524 \times 10^{-6}$ &\\
\hline
& GR ($l=3$) & 0 & $0.672188 - 0.0960977i$ & $0.4719 \times 10^{-3}$ &\\ \hline
&\multirow{4}{4em}{$l=3$} 
  & $1$ & $0.671556 - 0.096059 i$  & $0.0152 \times 10^{-5}$ &\\
&  & $5$ & $0.669059 - 0.095911i$  & $0.0351 \times 10^{-5}$ &\\
&  & $10$ & $0.666003 - 0.095734i$  & $0.5420 \times 10^{-5}$ &\\
&  & $15$ & $0.663016 - 0.095565i$  & $0.01269 \times 10^{-4}$ &\\
\hline\hline
\end{tabular}
\label{table1}
\end{table}
\begin{table}[!h]
\centering
\caption{QNMs of BHs described by the solution \eqref{eq46}, computed using 
the 6th-order Pad\'e-averaged WKB approximation method for different negative 
values of the parameter $\zeta$, with $M =1$, $n = 0$ and 
$\bar{\Lambda} = 0.001$.}
\vspace{5pt}
\begin{tabular}{c@{\hskip 5pt}c@{\hskip 10pt}c@{\hskip 10pt}c@{\hskip 10pt}c@{\hskip 5pt}c}
\hline\hline
& Multipole & $\zeta$ & Pad\'e averaged 6th-order WKB & $\Delta_{6}$ \\
\hline
&\multirow{4}{4em}{$l=1$} 
  & $-1$ & $0.291624 - 0.097502i$ & $0.1430 \times 10^{-5}$ &\\
&  & $-5$ & $0.292770 - 0.097625i$ & $0.0325 \times 10^{-5}$ &\\
&  & $-10$ & $0.294231 - 0.097789i $ & $0.0325 \times 10^{-5}$ &\\
&  & $-15$ & $0.295727 - 0.097963i$ & $0.1035 \times 10^{-4}$ &\\
\hline
&\multirow{4}{4em}{$l=2$} 
  & $-1$ & $0.481740 - 0.096430i$  & $0.0238 \times 10^{-4}$ &\\
&  & $-5$ & $0.483589 - 0.096592i$  & $0.5416 \times 10^{-5}$ &\\
&  & $-10$ & $0.485951 - 0.096806i$  & $0.0023 \times 10^{-5}$ &\\
&  & $-15$ & $0.488371 - 0.097034i$  & $0.0362 \times 10^{-5}$ &\\
\hline
&\multirow{4}{4em}{$l=3$} 
  & $-1$ & $0.672822 - 0.096136i$  & $0.0023 \times 10^{-5}$ &\\
&  & $-5$ & $0.675391 - 0.096293i$  & $0.4521 \times 10^{-5}$ &\\
&  & $-10$ & $0.678673 - 0.096501i$  & $0.0105 \times 10^{-5}$ &\\
&  & $-15$ & $0.682036 - 0.096721i$  & $0.3015 \times 10^{-4}$ &\\
\hline\hline
\end{tabular}
\label{table2}
\end{table}
Figure~\ref{fig2} illustrates the behaviour of the real and imaginary parts of 
the QNMs as functions of positive values of the parameter $\zeta$. The left 
panel shows the variation of the real part of the QNMs, while the right panel 
displays the variation of the imaginary part of the QNMs with respect to 
$\zeta$. It can be observed from the figure that both oscillation and damping 
decrease with the values of $\zeta$. Further, for negative values of the 
parameter $\zeta$, Fig.~\ref{fig3} shows the variation of the real and 
imaginary parts of the QNMs with respect to negative values of $\zeta$. In 
this case, the left panel indicates that the real part of the QNMs increases 
with increasing negative values of $\zeta$, while the right panel shows a 
corresponding increase in the imaginary part.
\subsection{QNMs in Schwarzschild-Anti-de Sitter spacetime} \label{sub2}
The SAdS spacetime describes a BH embedded in a background with a negative 
cosmological constant. In contrast to the SdS case, spacetime possesses 
only a BH event horizon, while the SAdS boundary at spatial infinity plays an 
important role in determining the global structure and dynamical properties of 
spacetime. In this section, we study the QNMs of BHs described by solution
\eqref{eq46} for a negative cosmological constant corresponding to the SAdS 
spacetime. In SAdS geometries, the perturbation equations possess regular 
singular points at both the event horizon and at spatial infinity. The 
tortoise coordinate $r_{\star}$ diverges to $-\infty$ as $r \to r_{h}$, where 
$r_h$ denotes the event horizon radius, while $r_{\star}$ approaches a finite 
constant as $r \to \infty$ \cite{adsqnm1}. Moreover, since in the SAdS 
spacetime the tortoise coordinate $r_{\star}$ approaches a finite constant as 
$r \to \infty$ \cite{adsqnm1}, consequently, the boundary conditions required 
for QNMs differ from the asymptotically flat or SdS spacetime. However, the 
condition at the event horizon remains the same: only purely ingoing modes 
are allowed, so that
\begin{equation}
\Psi(r_{\star}) \to A\, e^{i\omega r_{\star}}, \qquad r_{\star} \to -\infty .
\end{equation}
At spatial infinity, the scalar field perturbation Eq.~\eqref{eq50} 
yields the asymptotic behaviour as
\begin{equation}
\Psi \sim C r^{-2} + D r, \qquad r \to \infty .
\end{equation}
Here, $C$ and $D$ are two constant coefficients. Regularity of the scalar 
field at the SAdS boundary requires the absence of the divergent term, 
implying $D = 0$. This corresponds to imposing a Dirichlet boundary condition 
at infinity for determining the QNMs. 

\begin{figure}[!h]
\centerline{
\includegraphics[scale=0.87]{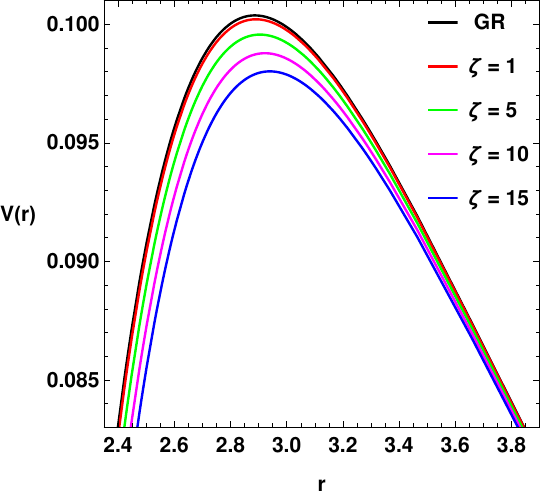} \hspace{0.7cm}
\includegraphics[scale=0.87]{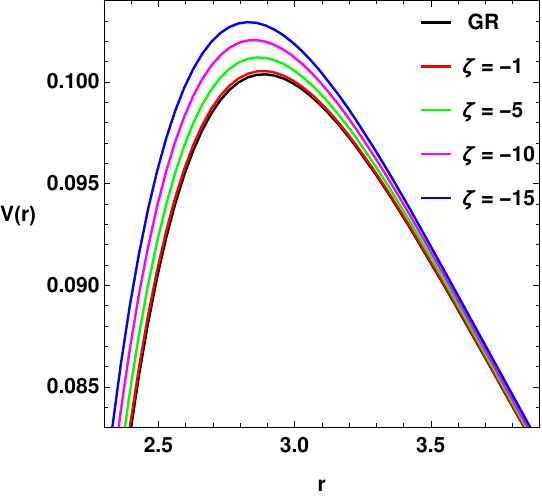}}
\vspace{-0.2cm}
\caption{Behaviour of the effective potential $V(r)$ as a function of the 
radial coordinate $r$ for both positive and negative values of $\zeta$, fixing 
$M = 1$, and $\bar{\Lambda} = -0.001$.} \label{fig4}
\end{figure}

Fig.~\ref{fig4} shows the variation of $V(r)$ with respect to $r$ for 
different values of $\zeta$ in SAdS BH spacetime. The left panel corresponds 
to positive values of $\zeta$. It is clear from the left panel that, for 
positive values of $\zeta$, the peak of the effective potential decreases as 
$\zeta$ increases. The right panel shows that, for negative values of $\zeta$, 
the height of the potential increases as negative values of $\zeta$ increase.

The WKB approximation cannot be reliably applied to SAdS spacetimes because 
the SAdS boundary at spatial infinity is reflective rather than purely 
outgoing. Therefore, in this section, we compute the QNMs using the direct 
shooting method described in Refs.~\cite{shooting1,shooting2,shooting3,
shooting4}. The radial equation is integrated outward from the horizon with 
the ingoing boundary condition. At large radius, the solution behaves as 
$\Psi(r)=Ar + B/r^{l+1}$, and the QNMs are obtained by imposing the Dirichlet 
boundary condition $A=0$. The uncertainty of the QNM frequencies is estimated 
from the numerical stability of the shooting procedure. For a fixed value of 
the parameter $\zeta$, the computation is repeated by varying the asymptotic 
radius, the integration domain, and the solver accuracy. The frequency is 
found to be stable up to six decimal places under these variations. Therefore, 
the results are used up to six decimal places.

Table.~\ref{table3} shows the variation of the QNMs with respect to both 
positive and negative values of $\zeta$ for different values of $l$. It can be 
observed from the table that for each $l$, both real and imaginary part of the QNMs increases for positive values of 
$\zeta$. On the other hand, for negative values of $\zeta$, both real and imaginary part of QNMs decreases with the increasing negative values 
of $\zeta$. Further, as $l$ increases, both real and imaginary parts increase 
monotonically. Moreover, similar to the SdS case, the variation of the QNMs 
with respect to $\zeta$ is small for low values of $\zeta$, and becomes more 
prominent for larger $\zeta$.
\begin{table}[!h]
\centering
\caption{QNMs of BHs described by the solution \eqref{eq46}, computed using 
the Direct Shooting Method for both positive and negative values of the 
parameter $\zeta$, with $n = 0$, $M =1$ and $\bar{\Lambda} = -0.001$.}
\vspace{5pt}
\begin{tabular}{c@{\hskip 5pt}c@{\hskip 10pt}c@{\hskip 10pt}c@{\hskip 10pt}|c@{\hskip 10pt}c@{\hskip 5pt}c}
\hline\hline
 & Multipole & $\zeta$ & Direct Shooting Method & $\zeta$ & Direct Shooting Method& \\
\hline

 & GR ($l=1$) & 0 & $0.381323 - 0.034391i$ & 0 & $0.381323 - 0.034391i$ \\
\hline

 & \multirow{4}{*}{$l=1$}
 & $1$  & $0.381359 - 0.034444i$ & $-1$  & $0.381287 - 0.034338i$ \\
 &    & $7$  & $0.381572 - 0.034757i$ & $-7$  & $0.381067 - 0.034014i$ \\
 &    & $14$ & $0.381814 - 0.035112i$ & $-14$ & $0.380805 - 0.033626i$ \\
 &    & $21$ & $0.382051 - 0.035458i$ & $-21$ & $0.380536 - 0.033224i$ \\
\hline

 & GR ($l=2$) & $0$ & $0.489696 - 0.038371i$ & $0$ & $0.489696 - 0.038371i$ \\
\hline

 & \multirow{4}{*}{$l=2$}
 & $1$  & $0.489727 - 0.038411i$ & $-1$  & $0.489666 - 0.038332i$ \\
 &    & $7$  & $0.489908 - 0.038645i$ & $-7$  & $0.489480 - 0.038089i$ \\
 &    & $14$ & $0.490115 - 0.038909i$ & $-14$ & $0.489258 - 0.037796i$ \\
 &    & $21$ & $0.490318 - 0.039166i$ & $-21$ & $0.489032 - 0.037492i$ \\
\hline

 & GR ($l=3$) & $0$ & $0.673089 - 0.045701i$ & $0$ & $0.673089 - 0.045701i$ \\
\hline

 & \multirow{4}{*}{$l=3$}
 & $1$  & $0.673131 - 0.045771i$ & $-1$  & $0.673046 - 0.045629i$ \\
 &    & $7$  & $0.673381 - 0.046190i$ & $-7$  & $0.672790 - 0.045194i$ \\
 &    & $14$ & $0.673666 - 0.046665i$ & $-14$ & $0.672484 - 0.044671i$ \\
 &    & $21$ & $0.673943 - 0.047126i$ & $-21$ & $0.672174 - 0.044129i$ \\
\hline\hline
\end{tabular} \label{table3}
\end{table} 

\begin{figure}[!h]
    \centerline{
    \includegraphics[scale = 0.86]{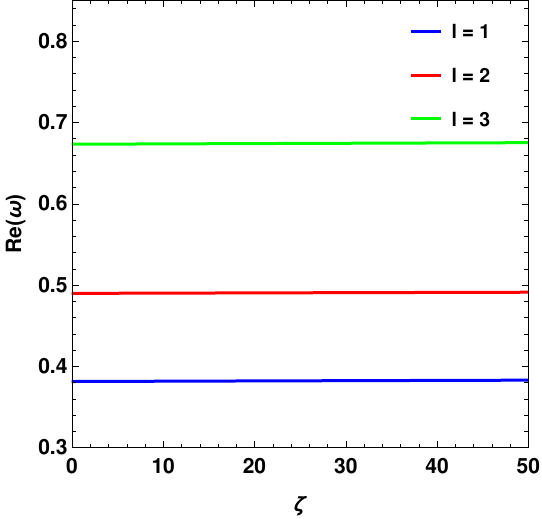}\hspace{0.5cm}
    \includegraphics[scale = 0.90]{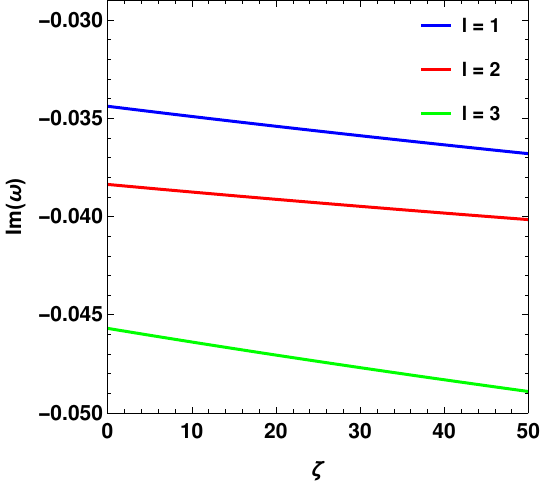}}
    \vspace{-0.2cm}
    \caption{Variation of real and imaginary parts of QNMs with respect 
to positive values of $\zeta$ for different values of $l$ along with $M = 1$, 
$n = 0$, and $\bar{\Lambda} = -0.001$.}
    \label{fig5}
\end{figure}

\begin{figure}[!h]
    \centerline{
    \includegraphics[scale = 0.86]{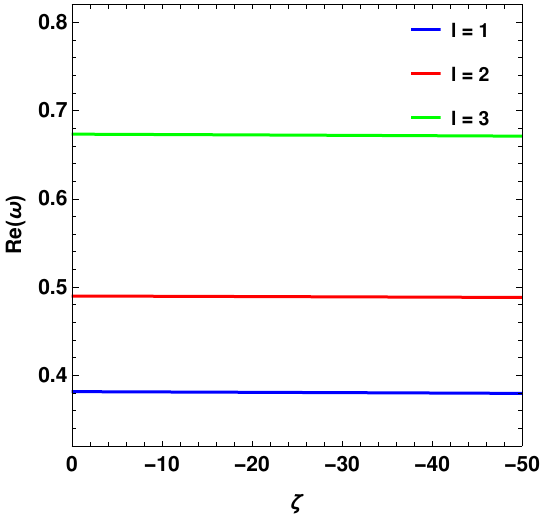}\hspace{0.5cm}
    \includegraphics[scale = 0.90]{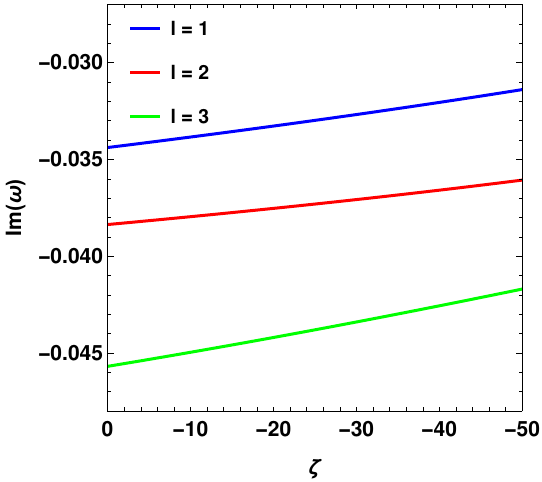}}
    \vspace{-0.2cm}
    \caption{Variation of real and imaginary parts of QNMs with respect 
to negative values of $\zeta$ for different values of $l$ along with $M = 1$, 
$n = 0$, and $\bar{\Lambda} = -0.001$.}
    \label{fig6}
\end{figure}

For visual realization, Fig.~\ref{fig5} shows the variation of the real and 
imaginary parts of the QNMs with respect to positive values of $\zeta$. It is 
clear from the figure that both real and the imaginary part 
increases as $\zeta$ increases. Similarly, Fig.~\ref{fig6} shows the variation 
of the QNMs for negative values of $\zeta$. As seen in Fig.~\ref{fig6}, the 
both real and the imaginary part decreases with the values of 
$\zeta$, for each $l$.
\section{Evolution of Scalar Perturbations around the Black Holes} \label{sec5}
In this section, we investigate the time evolution of scalar field 
perturbations in the BH spacetimes described by the solution~\eqref{eq46}, for 
both SdS and SAdS spacetimes. For this purpose, we employ the time-domain 
integration method described in Refs.~\cite{DJ2,Gundlach}. Accordingly, 
defining $\Psi(r_{\star},t) = \Psi(i\,\Delta r_{\star}, j\,\Delta t) = \Psi_{i,j}$ and $V(r(r_\star)) = V(r_{\star},t) = V_{i,j}$, 
Eq.~\eqref{eq50} can be written in the following form:
\begin{equation}
\frac{\Psi_{i+1,j}-2\Psi_{i,j}+\Psi_{i-1,j}}{{\Delta}r_{\star}^2}-\frac{\Psi_{i,j+1}-2\Psi_{i,j}+\Psi_{i,j-1}}{{\Delta}t^2}-V_{i}\Psi_{i,j}=0. 
\label{eq56}
\end{equation}
Now, using the initial conditions: $\Psi(r_{\star},t) = 
\exp\left[-(r_{\star}-q)^2/2\sigma^2\right]$ and 
$\left|\Psi(r_{\star},t)\right|_{t<0} = 0$, where $q$ is the median and 
$\sigma$ is the width of the initial wave packet, the time evolution of the 
scalar field can finally be written as
\begin{equation}
\Psi_{i,j+1} = -\,\Psi_{i,j-1}+\left(\frac{{\Delta}t}{{\Delta}r_{\star}}\right)^{\!2}\left(\Psi_{i+1,j}+\Psi_{i-1,j}\right)+\left(2-2\left(\frac{{\Delta}t}{{\Delta}r_{\star}}\right)^{\!2}-V_{i\,}\Delta{t}^2\right)\Psi_{i,j}. 
\label{eq57}
\end{equation} 
Moreover, during the numerical procedure, we impose the Von Neumann stability 
condition $\frac{{\Delta}t}{{\Delta}r_{\star}}<1$ to ensure stable results 
and compute the time profiles.
\begin{figure}[!h]
    \centerline{
    \includegraphics[scale = 0.45]{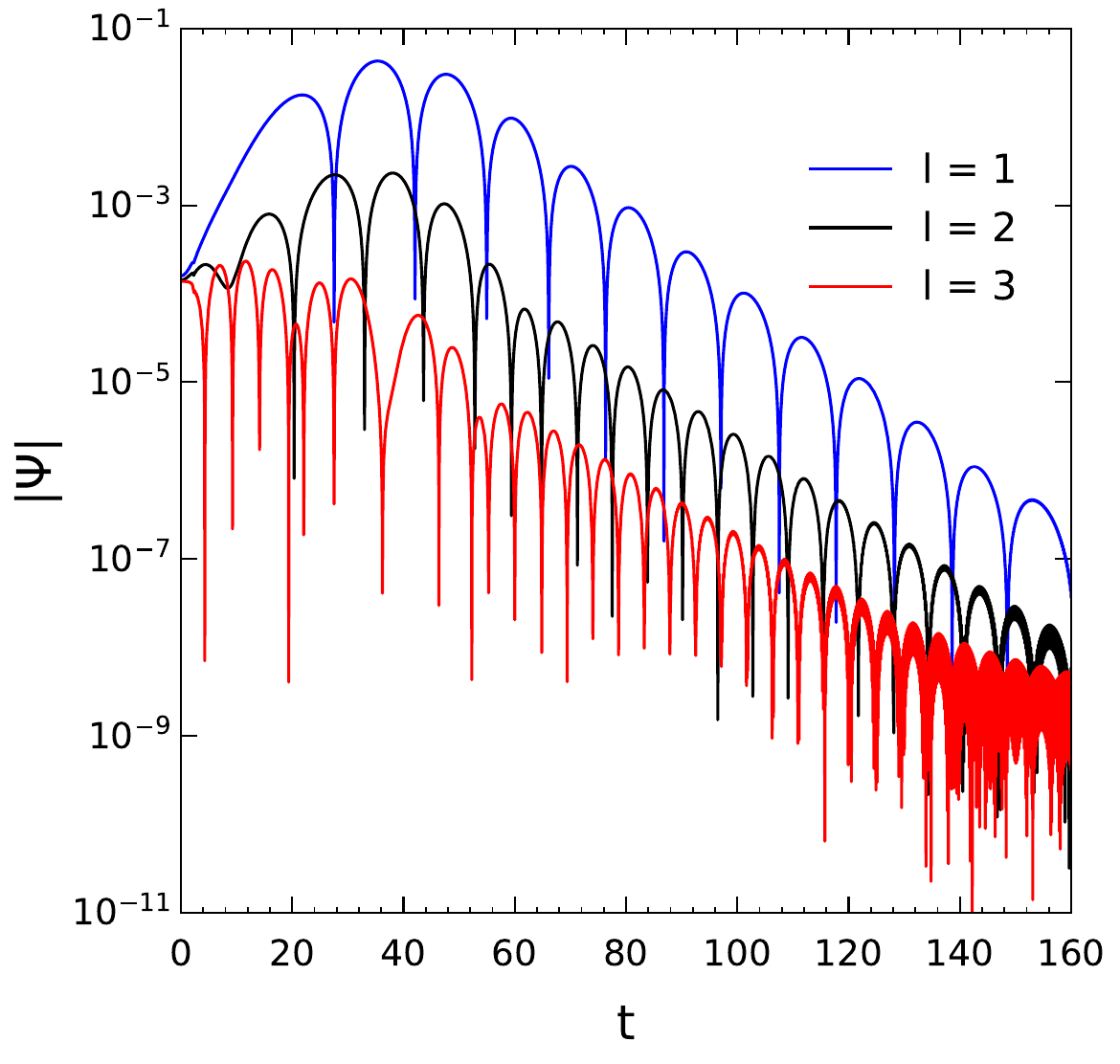}\hspace{0.5cm}
    \includegraphics[scale = 0.45]{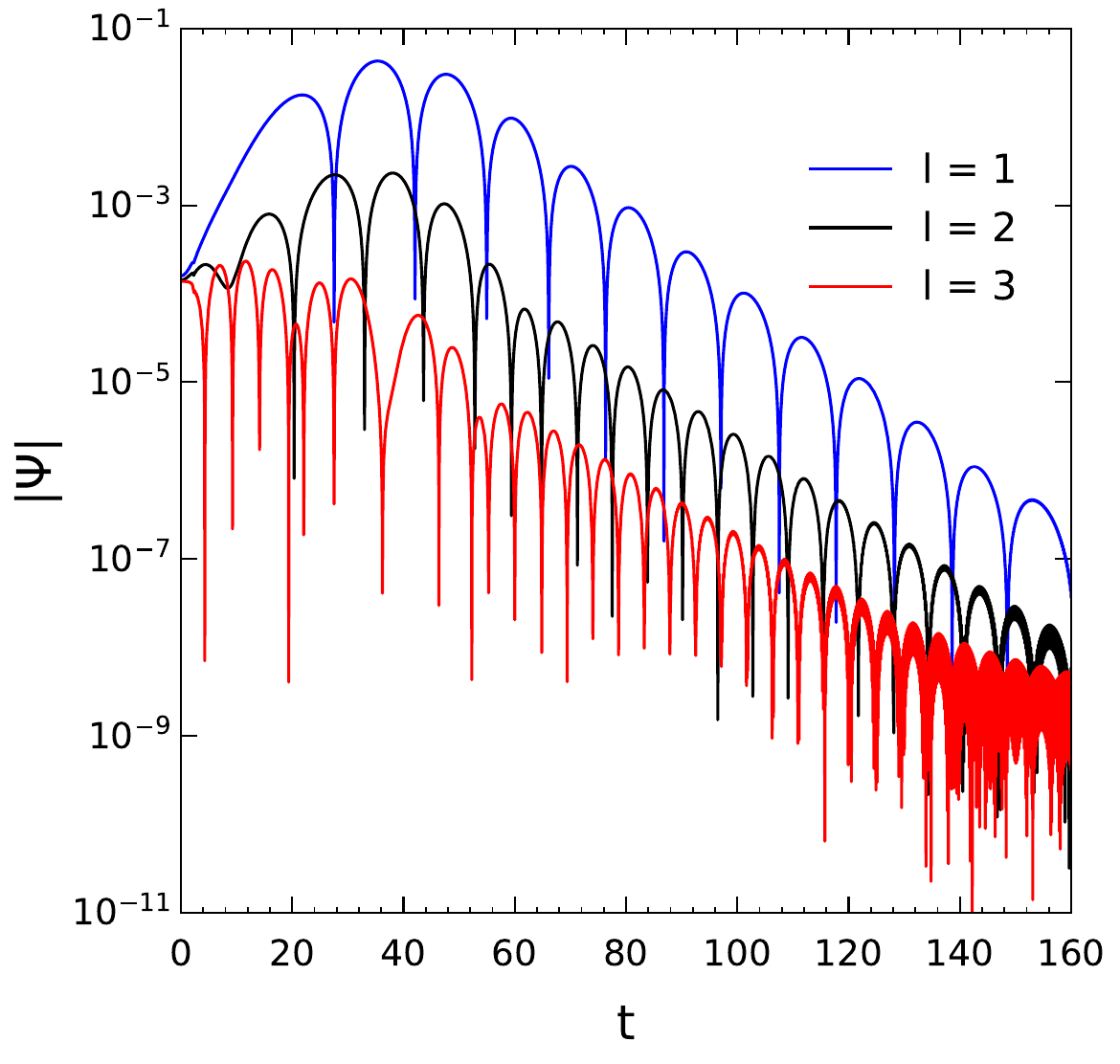}}
    \vspace{-0.2cm}
    \caption{Time-domain profiles of scalar field perturbations in the BH 
spacetime described by the metric \eqref{eq46} for the SdS case with 
$\zeta = 10$ (left plot) and  $\zeta = -10$ (right plot). For both
plots $\bar{\Lambda}=0.001$ and $n=0$ are used.}
    \label{fig7}
\end{figure}
Fig.~\ref{fig7} illustrates the time-domain profiles of scalar field 
perturbations in the BH spacetime for the SdS case, where the left panel 
corresponds to $\zeta = 10$, while the right panel corresponds to 
$\zeta = -10$. The dependence of the oscillation frequency and damping rate 
of the QNMs on $l$, discussed in Table~\ref{table1} and Table~\ref{table2}, is 
also reflected in Fig.~\ref{fig7}.
\begin{figure}[!h]
    \centerline{
    \includegraphics[scale = 0.45]{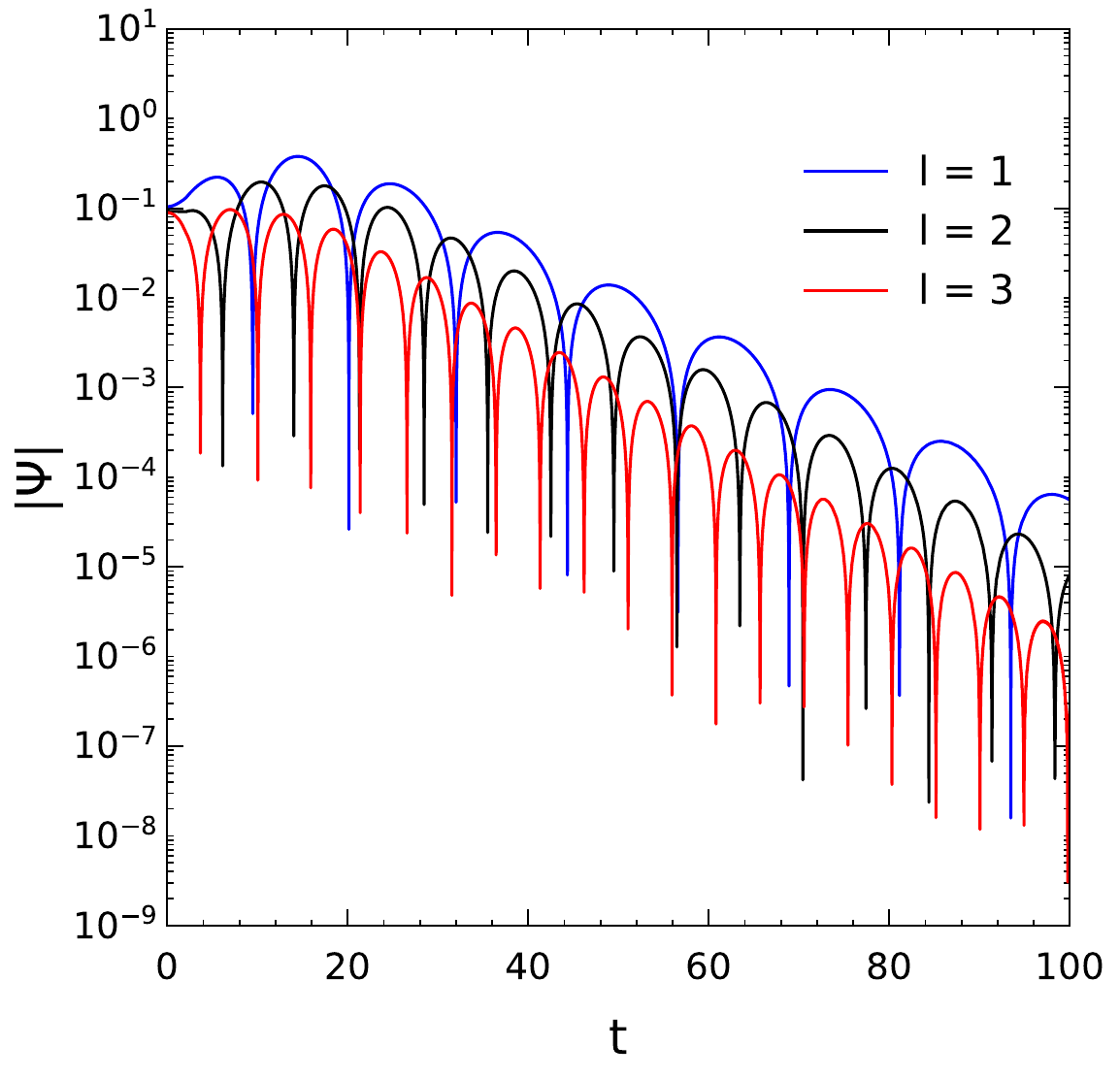}\hspace{0.5cm}
    \includegraphics[scale = 0.45]{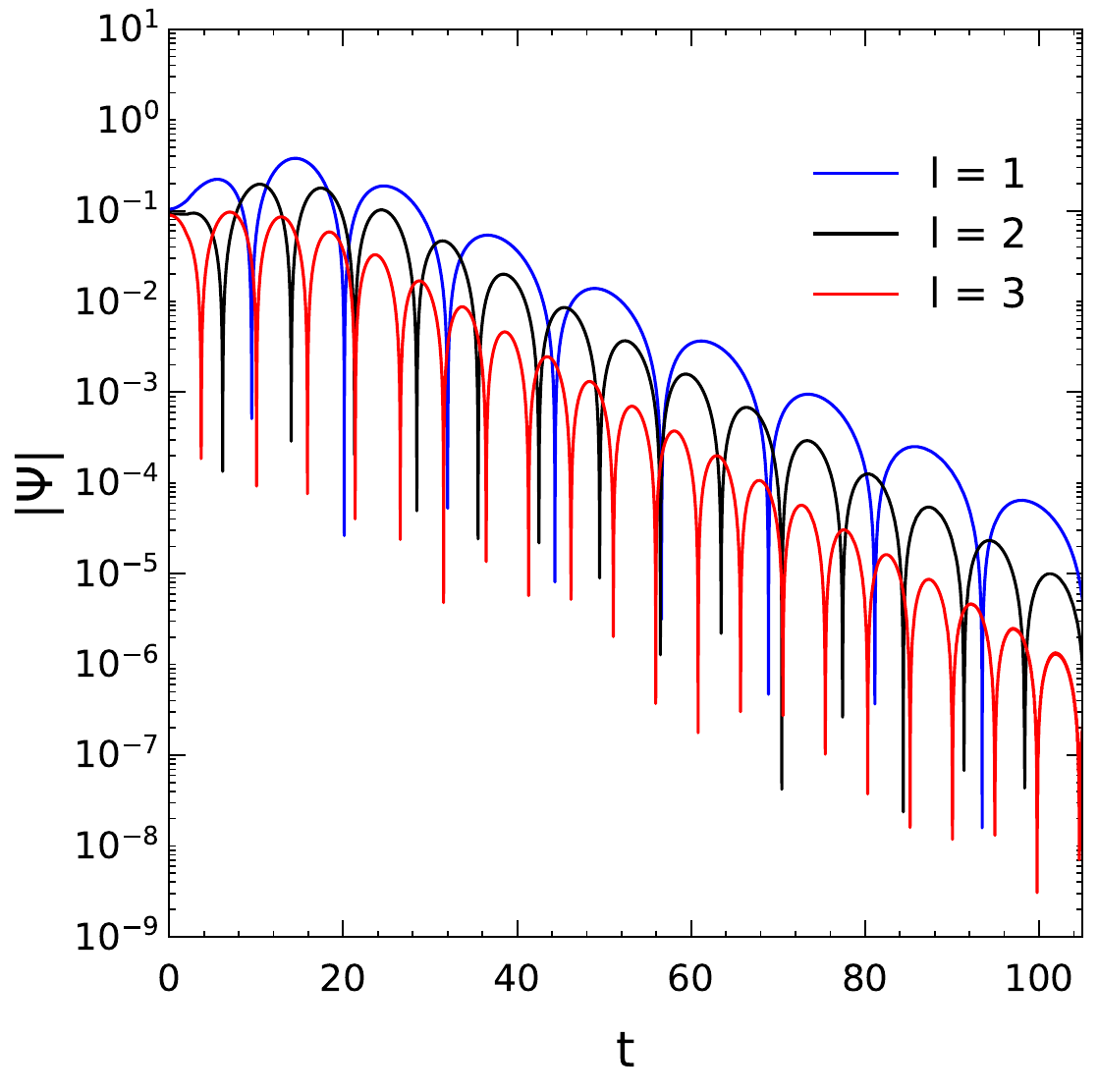}}
    \vspace{-0.2cm}
    \caption{Time-domain profiles of scalar field perturbations in the BH 
spacetime described by the metric \eqref{eq46} for the SAdS case with
$\zeta = 10$ (left plot) and  $\zeta = -10$ (right plot). For both
plots $\bar{\Lambda}=-0.001$, and $n=0$ are used.}
    \label{fig8}
\end{figure}
Fig.~\ref{fig8} shows the time-domain profiles of the scalar field 
perturbations in the SAdS spacetime, where the left panel corresponds to 
$\zeta = 10$, while the right panel corresponds to $\zeta = -10$. Similar to 
the SdS case, the dependence of the oscillation and damping behaviour on $l$ 
is also reflected in the time-domain profile.

Furthermore, we employed the Matrix Pencil (MP) method, as described in 
Refs.~\cite{MP1,MP2,MP3}, to extract the QNMs from the time-domain profiles. 
For this, from the time profile waveform, first we constructed two time 
shifted Hankel matrices $Y_0$ and $Y_1$. Then a singular value decomposition 
(SVD) was performed on $Y_1$ as $Y_1 = U S V^\dagger$, where $U$ and $V$ are 
unitary matrices containing the left and right singular vectors, respectively, 
and $S$ is a diagonal matrix whose diagonal entries are the singular values. 
We kept the largest singular values (say $p$) corresponding to the physical 
QNMs and discarded the remaining noise dominated components. The reduced 
matrices were then used to form the matrix $Z_p = S_p^{-1}U_p^\dagger Y_0 V_p$,
here $U_p$ and $V_p$ denote the matrices formed by the first $p$ columns of 
the left and right singular vector matrices $U$ and $V$, respectively, while 
$S_p^{-1}$ is the inverse of the diagonal matrix $S_p$ containing the $p$ 
largest singular values. The eigenvalues of $Z_p$ provide estimates of the 
inverse poles of the waveform. From these poles, we obtained the complex 
frequencies as described in \cite{MP3}, where the real part corresponds to 
the oscillation frequency and the imaginary part determines the damping rate. 
Finally, the extracted QNMs were used to reconstruct the time-domain waveform 
and matched against the original time-domain profile. The coefficient of 
determination $R^2$ is computed as
\begin{equation}
R^2 = 1 - \frac{\sum\left(\psi_{\text{Time-Profile}} - \psi_{\text{MP}}\right)^2}
{\sum\left(\psi_{\text{Time-Profile}} - \bar{\psi}_{\text{Time-Profile}}\right)^2},
\end{equation}
where $\psi_{\text{Time-Profile}}$ denotes the amplitude of the original 
time-domain waveform, $\psi_{\text{MP}}$ is the amplitude of the waveform 
reconstructed using the MP method, and $\bar{\psi}_{\text{Time-Profile}}$ is 
the mean value of the original time-domain waveform.

\begin{figure}[!h]
    \centerline{
    \includegraphics[scale = 0.4]{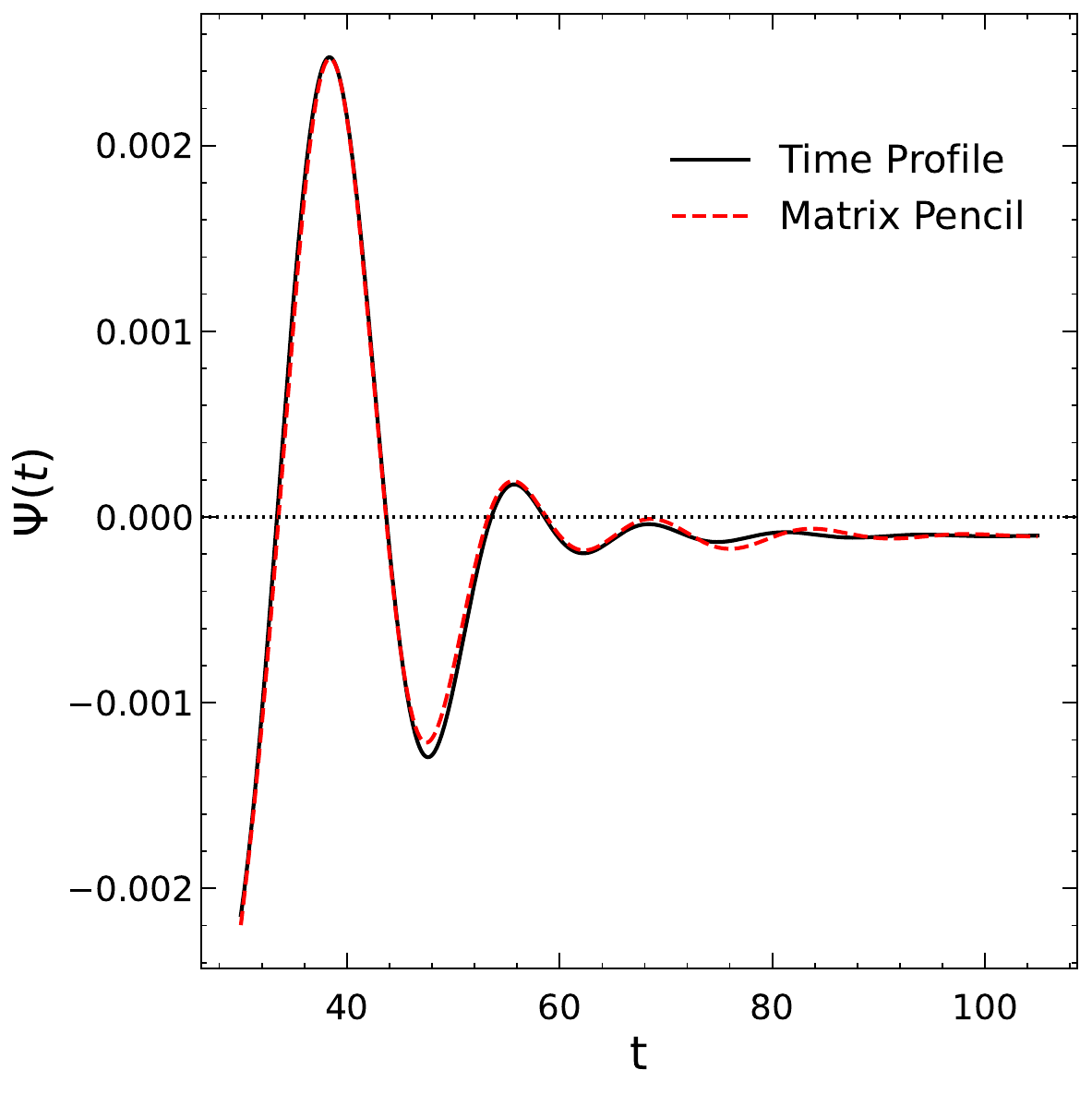}\hspace{0.5cm}
    \includegraphics[scale = 0.4]{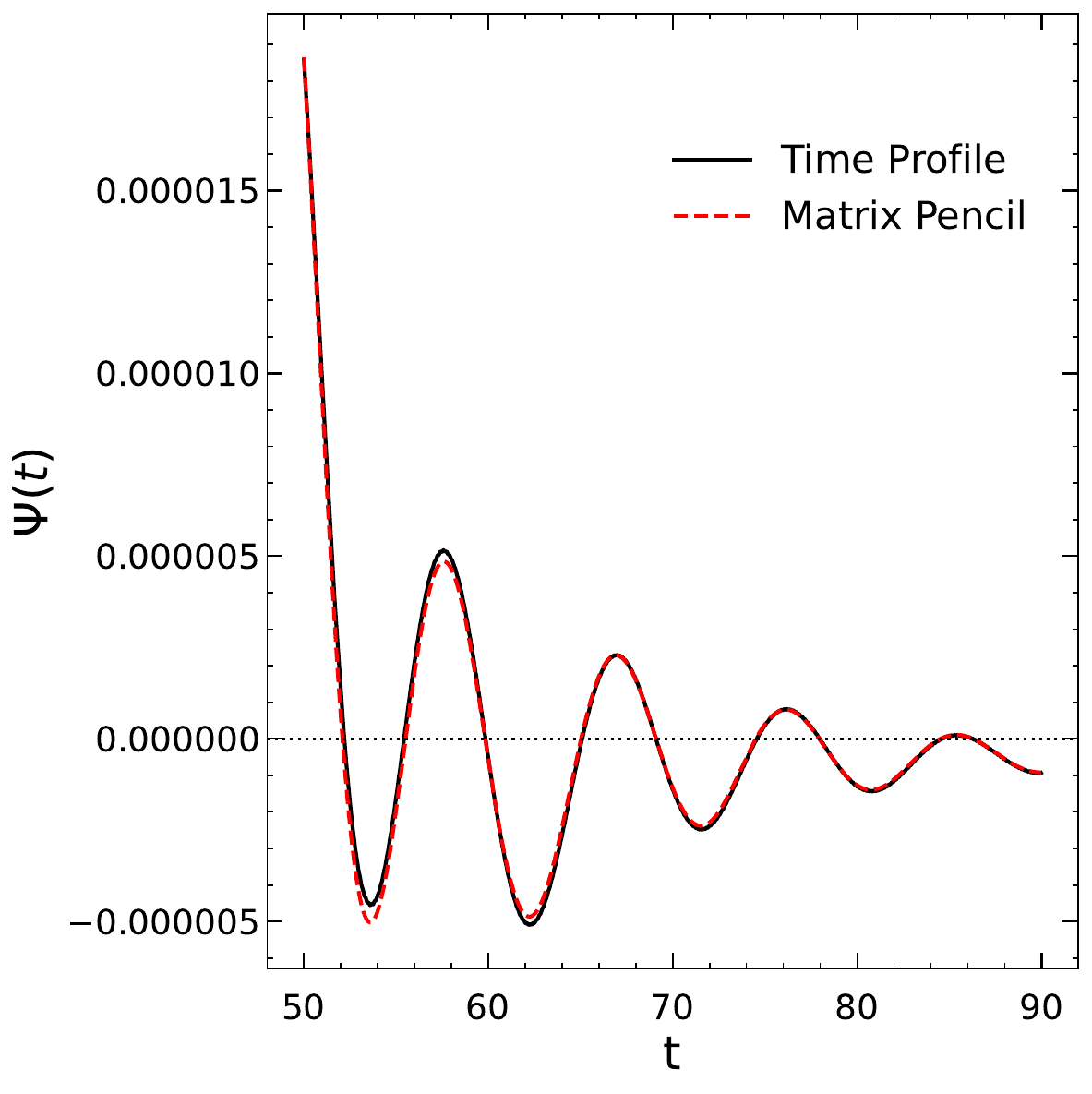}}
    \vspace{-0.2cm}
    \caption{Comparison between the scalar field QNM time-domain profiles and 
the waveforms reconstructed using the Matrix Pencil method for the SdS case 
with $\bar{\Lambda} = 0.001$, and $n = 0$. The left plot is for $l = 2$ with 
$\zeta = 10$, while the right plot is for $l = 3$ with $\zeta = -10$.
}
\label{fig9}
\end{figure}

\begin{figure}[!h]
    \centerline{
    \includegraphics[scale = 0.4]{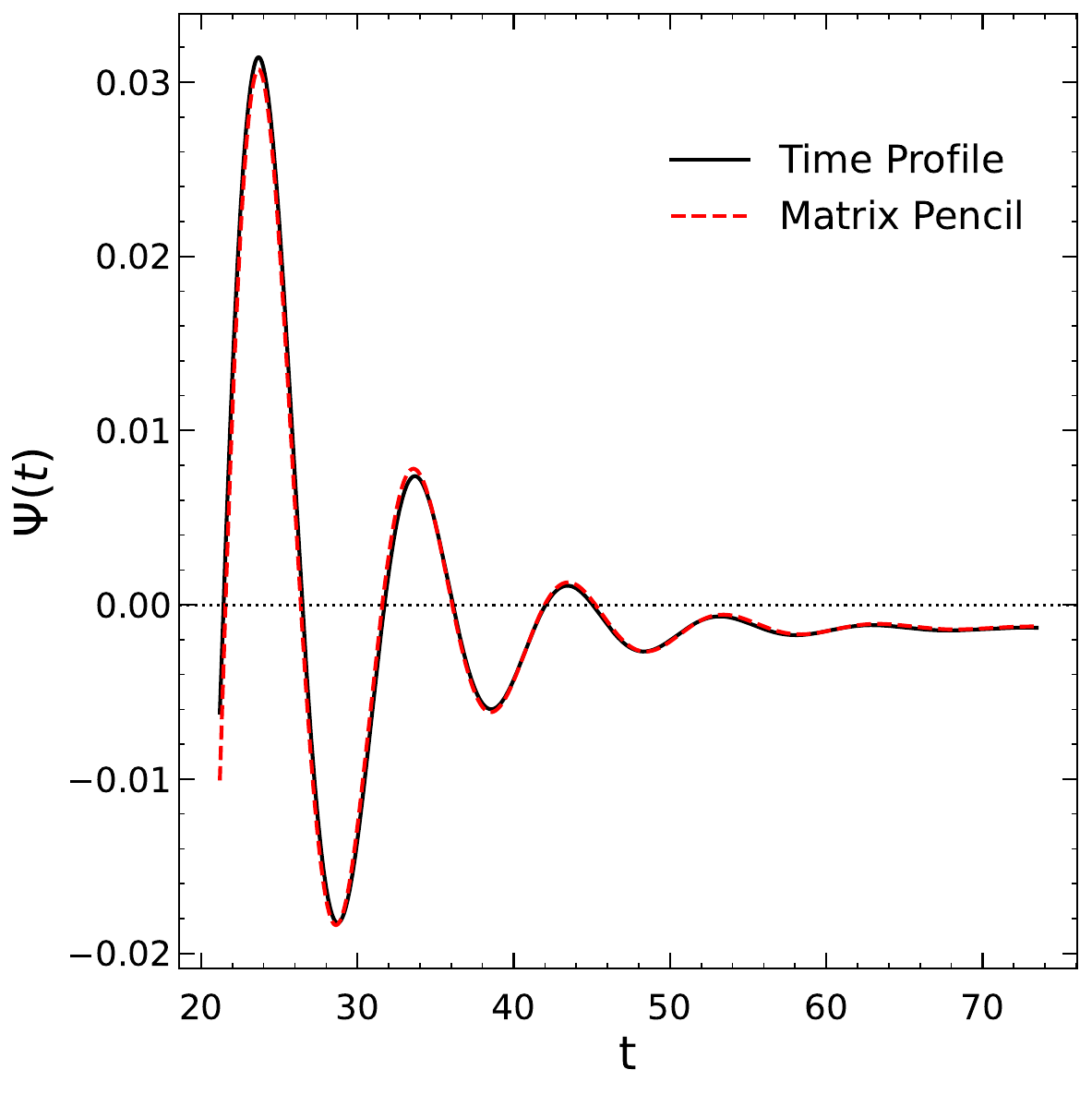}
    \hspace{0.5cm}
    \includegraphics[scale = 0.4]{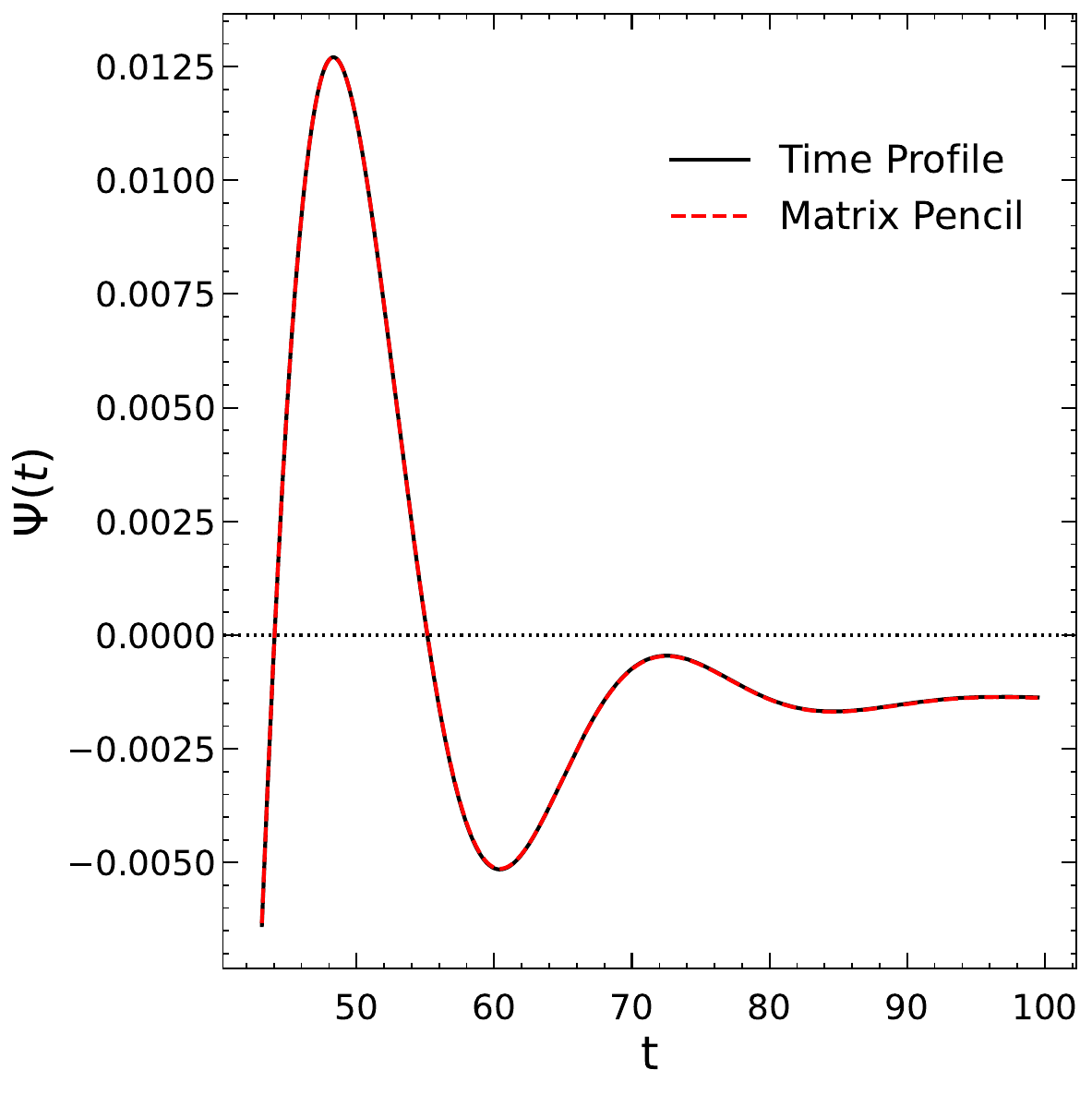}}
    \vspace{-0.2cm}
    \caption{Comparison between the QNM time-domain profile and the waveform 
reconstructed using the Matrix Pencil method for the SAdS case with 
$\bar{\Lambda} = -0.001$, $n = 0$. The left plot is for $l = 3$ with $\zeta = 10$ and the right plot is for $l = 1$ with $\zeta = -10$.
}
    \label{fig10}
\end{figure}

Fig.~\ref{fig9} shows a comparison between the waveforms reconstructed using 
the MP method and the original time-domain profiles for the SdS case. The left 
panel corresponds to $l = 2$ with $\zeta = 10$, while the right panel 
corresponds to $l = 3$ with $\zeta = -10$. It can be seen that the waveform 
reconstructed from the QNMs extracted via the MP method is in good agreement 
with the original time-domain waveforms.
Table~\ref{tab4} presents a comparison between the QNMs extracted from the 
time-domain profile using the MP method and those obtained using the 
Pad\'e-averaged WKB method for the SdS case. It is evident that the QNMs 
derived from the time profile are in good agreement with the results from the 
Pad\'e-averaged WKB approach. We calculate the difference in the magnitudes of 
the QNMs obtained from the 6th-order Pad\'e averaged WKB approximation method 
and those obtained from the time domain profile using the MP method as follows:
\begin{equation}
\Delta_\text{QNM} = \frac{\left| \text{QNM}_{\text{WKB}} - \text{QNM}_{\text{Time-Profile}} \right|}{2}. 
\label{eq59}
\end{equation}
\begin{table}[!h]
\caption{Comparison of QNMs extracted from the time-domain profiles using the
Matrix Pencil method and that obtained from the Pad\'e averaged WKB method.}
\vspace{5pt}
    \centering
    \begin{tabular}{|@{\hskip 5pt}c@{\hskip 5pt}| @{\hskip 5pt}c@{\hskip 5pt}|@{\hskip 5pt}c@{\hskip 5pt}|@{\hskip 5pt}c@{\hskip 5pt}|@{\hskip 5pt}c@{\hskip 5pt}|@{\hskip 5pt}c@{\hskip 5pt}|}
        \hline
        \textit{l} & $\zeta$ &Time-domain & $R^2$ & Pad\'e averaged WKB & $\Delta_\text{QNM}$ \\[2pt]
        \hline
        $l = 1$ & $3$ &$0.291762 - 0.097415i$ & $0.995360$ & $0.290497 - 0.097385i$ & $0.000633$ \\
        $l = 2$ & $10$ &$0.472973 - 0.096106$ & $0.997538$ & $0.476832 - 0.096018i$ & $0.001930$\\
        $l = 3$ & $16$ & $0.669175 - 0.095286i$ & $0.998405$ & $0.662426 - 0.095532i$ & $0.003377$ \\
        $l = 1$ & $-3$ &$0.294631 - 0.097917i$ & $0.987225$ & $0.292194 - 0.097563i$ & $0.001231$ \\
        $l = 2$ & $-10$ &$0.481760 - 0.096781i$ & $0.9955785$ & $0.485951 - 0.096806i$ & $0.002096$\\
        $l = 3$ & $-16$ & $0.685193 - 0.096542i$ & $0.999740$ & $0.682719 - 0.096767i$ & $0.001242$ \\
        \hline
    \end{tabular}
    \label{tab4}
\end{table}
Moreover, Fig.~\ref{fig10} illustrates a comparison between the waveforms
of time-domain profiles and those obtained from the MP method 
for the SAdS case. The left panel corresponds to $l = 3$ with $\zeta = 10$ and right panel corresponds to $l = 1$ with $\zeta = -10$. The 
figure clearly shows that, for the SAdS case also the two waveforms are in 
good agreement.
Table~\ref{tab5} presents the QNMs obtained from the direct shooting method 
and those extracted from the time-domain profile for the SAdS case. The close 
agreement between the two sets of results demonstrates the reliability of our 
analysis. The difference between the QNMs obtained from the time-domain method and those obtained from the direct shooting method is calculated using
\begin{equation}
\Delta_\text{QNM} = \frac{\left| \text{QNM}_{\text{Direct Shooting}} - \text{QNM}_{\text{Time-Profile}} \right|}{2}. 
\end{equation}
\begin{table}[!h]
\caption{Comparison of QNMs extracted from the time-domain profiles using 
the Matrix Pencil method and that obtained from the Direct Shooting method, 
for the SAdS case.}
\vspace{5pt}
    \centering
    \begin{tabular}{|@{\hskip 5pt}c@{\hskip 5pt}| @{\hskip 5pt}c@{\hskip 5pt}|@{\hskip 5pt}c@{\hskip 5pt}|@{\hskip 5pt}c@{\hskip 5pt}|@{\hskip 5pt}c@{\hskip 5pt}|@{\hskip 5pt}c@{\hskip 5pt}|}
        \hline
        \textit{l} & $\zeta$ &Time domain & $R^2$ & Direct Shooting method & $\Delta_\text{QNM}$ \\[2pt]
        \hline
        $l = 1$ & $3$ &$0.388571 - 0.034287i$ & $0.999360$ & $0.381430 - 0.034549i$ & $0.003573$ \\
        $l = 2$ & $10$ &$0.481796 - 0.038596i$ & $0.991538$ & $0.489997 - 0.038759i$ & $0.004101$\\
        $l = 3$ & $16$ & $0.673482 - 0.046128i$ & $0.999230$ &$0.673746 - 0.046798i$ & $0.000360$ \\
        $l = 1$ & $-3$ &$0.383782 - 0.034487i$ & $0.998722$ & $0.381214 - 0.034231i$ & $0.001290$ \\
        $l = 2$ & $-10$ &$0.489247 - 0.037982i$ & $0.999982$ & $0.489385 - 0.037965i$ & $0.002571$\\
        $l = 3$ & $-16$ & $0.675792 - 0.044612i$ & $0.997250$ & $0.672396 - 0.044518i$ & $0.001699$ \\
        \hline
    \end{tabular}
    \label{tab5}
\end{table}

In the present work, we restrict our analysis to scalar field 
perturbations as scalar field perturbations provide a comparatively simpler 
framework for probing the properties of BH spacetimes and 
their associated QNMs. As the effective potential governing scalar 
perturbations depends on the background geometry, scalar QNMs can effectively 
capture the influence of the RG-improved corrections on the spacetime. 
Further, although the quantitative spectra of electromagnetic and gravitational 
perturbations may differ from the scalar field case, it is expected 
to have similar qualitative behaviour, because the corresponding perturbation 
equations also depend on the underlying BH geometry.

It can be observed from the above analysis that the results obtained from 
different computational methods are in very good agreement. The present 
RG-improved corrections are perturbatively small, and the corresponding 
shifts in the QNM frequencies are also small. However, they can influence the 
photon sphere structure and also the shadow radius. The presence of the 
additional term in the metric modifies the QNM frequencies, which leads to 
affecting the ringdown signal of GWs and also the changes in the near BH 
geometry can alter BH shadow observables. Therefore, high-precision 
observations from future GW detectors such as the Laser Interferometer 
Space Antenna (LISA) \cite{LISA1,LISA2}, a joint ESA--NASA mission, 
TianQin \cite{TianQin1,TianQin2} and Taiji \cite{Taiji1,Taiji2} proposed 
by China, the DECi-hertz Interferometer Gravitational wave Observatory 
(DECIGO) \cite{japan1,japan2} proposed by Japan, and LIGO--India 
\cite{ILIGO1,ILIGO2}, an Indo--US collaboration, together with future BH 
imaging experiments such as the next-generation Event Horizon Telescope 
(ngEHT) \cite{ngEHT1,ngEHT2}, an international collaboration, and the 
Black Hole Explorer (BHEX) mission \cite{BHEX1,BHEX2}, a proposed USA-led 
international mission, may provide possible indications of the presence 
of the additional term in the metric and could offer phenomenological 
support for RG-improved black hole spacetimes. 
Similar higher-order corrections to 
the classical BH solution also arise in various EFTs and 
QG-inspired approaches \cite{HOC1,HOC2,HOC3,HOC4}.

\section{Summary and Conclusion} \label{sec6}
The goal of this work is to construct an RG improved perturbative BH solution, 
and to study the QNMs of the corresponding BHs in both SdS and SAdS 
spacetimes. The BH solution constructed in this work is an EFT solution, in 
which low-energy corrections are incorporated into the GR sector. These 
corrections can encode low-energy QG effects. Studying the QNMs provides 
insight into the impact of such tiny corrections and possible QG effects.

In Section \ref{sec2}, we briefly discuss the RG improved action and the 
corresponding field equations derived from it, following Refs.~\cite{RG1,RG2}. 
The BH solution is constructed in Section \ref{sec3} up to 2nd-order 
corrections. Since the correction parameter $\epsilon \ll 1$, we truncate the 
solution at second order as higher-order terms involve higher powers of 
$\epsilon$ and are therefore suppressed.

The QNMs of the BHs are investigated in Section \ref{sec4}. In this work, we 
consider a scalar field perturbation and analyze the QNMs for both SdS and 
SAdS spacetimes. In Subsection \ref{sub1}, we focus on the SdS spacetime. The 
behaviour of the reduced potential is examined for different values of $\zeta$ 
with $k_0 = 10$, $M = 1$, and $\bar{\Lambda} = 0.001$. It is found that, for 
positive values of $\zeta$, the peak of the potential decreases relative to 
the GR case, whereas for negative values of $\zeta$ the peak  increases compared to the GR case. The behaviour of the reduced potential with respect 
to $r$ for different values of $\zeta$ is shown in Figure \ref{fig1}. The QNMs 
are computed using the 6th-order Pad\'e-averaged WKB approximation for 
different values of $\zeta$. For positive values of $\zeta$, both the real and 
imaginary parts of the QNMs decrease, whereas for negative values of $\zeta$ 
they increase. Moreover, for both positive and negative $\zeta$, the real parts increases and imaginary parts decreases with the multipole number $l$. The behaviour of real 
and imaginary parts of QNMs with respect to $\zeta$ and $l$ is shown in Figures
\ref{fig2} and \ref{fig3} and tabulated in Tables \ref{table1} and 
\ref{table2}. In subsection \ref{sub2}, we study the QNMs in the SAdS 
spacetime using the direct shooting method. The behaviour of the effective 
potential $V(r)$ in this case as a function of $r$ for different values of 
$\zeta$ is shown in Figure \ref{fig4}. The corresponding QNMs are listed in 
Table \ref{table3} for both positive and negative values of $\zeta$ and for 
different multipole numbers $l$. Figure \ref{fig5} illustrates the variation 
of real and imaginary parts of QNMs with respect to positive values of $\zeta$, 
and Figure \ref{fig6} shows the variation of real and imaginary parts of 
QNMs with respect to negative values of $\zeta$ in SAdS spacetime.

Further, the time evolution of a scalar perturbations is studied in Section 
\ref{fig5} for both SdS and SAdS spacetimes. The time-domain profile for the 
scalar field perturbations is studied using Eq.~\eqref{eq57}. 
Figure~\ref{fig7} displays the time-domain profiles of the scalar field 
perturbations for the SdS case, and Figure~\ref{fig8} shows the time-domain 
profiles for the SAdS case. The behaviour of oscillation and damping of QNMs 
with respect to $l$ reflected in the time-domain profile is in good agreement 
with the Pad\'e-averaged WKB approximation method for the SdS case and direct 
shooting 
method for SAdS case. To extract the QNMs from the time profile, we employed 
the MP method. The extracted QNMs are used to reconstruct the waveforms. In 
Figure~\ref{fig9} we show the comparison between the original time profile 
waveforms and the reconstructed waveforms using the QNMs obtained from the MP 
method, for the SdS case considering $l = 2$ and $l = 3$. Moreover, 
Table~\ref{tab4} illustrates the QNMs obtained from time-domain analysis and 
Pad\'e-averaged WKB approximation. It can be observed from the table that the 
QNMs obtained from both methods are in good agreement. Next, we extend this 
analysis to the SAdS case. Using the same MP method, the QNMs are extracted. 
The comparison between the extracted waveforms and the original time-domain 
waveforms is shown in Figure~\ref{fig10} for $l = 3$ and $l = 1$. Table~\ref{tab5} 
presents the QNMs obtained from the time-domain analysis and the direct 
shooting method for different values of $l$ and $\zeta$. For the SAdS case 
also the QNMs obtained from both methods are in good agreement.

The parameters of the proposed BH solution can be constrained through observable quantities associated with strong-field BH phenomena. 
The deviations in the QNM frequencies and damping lead to small modifications in the ringdown spectrum of GWs, and the effects in the near-BH geometry can modify the photon sphere structure and consequently the BH shadow radius. 
Since the QNM spectrum and BH shadow radius depend explicitly on the parameters of the solution, future high-precision GW observations and BH imaging experiments can
allow one to test or constrain the proposed model as discussed earlier.

Different RG-improved BH solutions have previously been investigated 
using different frameworks \cite{com1,com2,com3,com4,com5,com6}. In most of 
these studies, the RG improvement is implemented by introducing running 
gravitational couplings directly into the classical BH metric, which leads to 
a nonperturbative RG-improved BH solution. Such analyses have revealed modified 
horizon structures, corrections to thermodynamic properties of BHs, modified 
QNMs spectra, etc. Further, some recent studies have also obtained inverse 
power corrections to the Schwarzschild metric through RG-improved running 
couplings \cite{com4}. In contrast, the present work derives the BH solution 
perturbatively from the RG-improved field equations in a framework where the 
cosmological constant is treated as a scale dependent quantity. Consequently, 
the corrections obtained here appear as leading order perturbative 
modifications to the classical Schwarzschild-(A)dS geometry. Moreover, despite 
these differences in the theoretical constructions, all approaches indicate 
that RG-improved quantum corrections can lead to observable modifications in 
strong-field BH phenomena, including changes in the effective potential and 
the corresponding QNM spectrum. Similar to several previous RG-improved BH 
solutions, the present model predicts deviations from the classical 
Schwarzschild geometry in the strong-field regime and induces shifts in the 
QNM spectrum. 

Our study primarily focuses on the effects on QNMs arising from small 
higher-order corrections to GR due to RG improvement. Such corrections, 
obtained through RG improvement, can encode tiny QG effects. Studying BH 
observables such as QNMs can provide a probe to test these QG effects. 
The present work can be extended in several directions. In 
particular, it would be interesting to investigate electromagnetic and 
gravitational perturbations of the RG-improved BH solution and compare 
their QNM spectra with the scalar case considered here. Since the RG-
induced corrections modify the near BH geometry, further studies of BH 
shadows, photon sphere properties, and gravitational lensing can provide 
additional insight into the model. Moreover, extensions to rotating BH spacetimes 
and the study of Quasi Periodic Oscillations (QPOs) can provide a deeper 
understanding of RG-improved QG effects. In recent years, detailed studies of QPOs in various spherically symmetric BH spacetimes, including Sen BHs \cite{1qpo}, regular BHs \cite{2qpo}, modified gravity motivated geometries \cite{3qpo}, and BH solution with Gauss-Bonnet trace anomaly \cite{OB3} have demonstrated the potential of QPO observations to constrain BH parameters and probe deviations from classical GR. Further, investigating the impact of 
these corrections on GW ringdown modes and constraining the parameters of 
the model using future high-precision observations would be an interesting 
future direction of the present work. Overall, this study contributes to 
the understanding of low-energy QG effects and offers valuable insights for 
future research. 
\section*{Acknowledgements} UDG is thankful to the Inter-University Centre
for Astronomy and Astrophysics (IUCAA), Pune, India for the Visiting
Associateship of the institute.



\end{document}